\RequirePackage{ifpdf}
\documentclass[11pt,letterpaper]{article}
\pdfoutput=1
\usepackage{jheppub}
\usepackage{amsmath,amssymb,amsfonts,amscd}
\usepackage{graphicx}
\usepackage{subcaption}
\usepackage{verbatim}
\usepackage{fancybox}
\usepackage{changepage}
\usepackage{slashed}
\usepackage{url}
\usepackage{array}
\usepackage{hhline}
\usepackage{tabularx}
\usepackage{tikz}

\usepackage{rotating,booktabs}
\usepackage[verbose]{placeins}
\usepackage{blindtext}

\graphicspath {{figures/}}

\newcommand{\bse}{\begin{subequations}}
\newcommand{\ese}{\end{subequations}}
\newcommand{\be}{\begin{equation}\begin{aligned}}
\newcommand{\ee}{\end{aligned}\end{equation}}

\newcommand{\td}[1]{{\tilde #1}}

\newcommand{\cof}{{\rm cof}}

\newcommand{\Tr}{{\rm Tr}}
\newcommand{\pbra}[1]{\left(#1\right)}
\newcommand{\bbra}[1]{\left[#1\right]}

\newcommand{\dl}[1]{\underline{#1}}
\newcolumntype{x}[1]{>{\centering\arraybackslash}p{#1}}

\def\Tr{{\rm Tr \,}}

\def\w{\omega}

\def\tilde{\widetilde}
\def\hat{\widehat}
\def\bar{\overline}

\def\CF{{\mathcal F}}

\def\CN{{\mathcal N}}
\def\CO{{\mathcal O}}

\def\CR{{\mathcal R}}
\def\CS{{\mathcal S}}
\def\CT{{\mathcal T}}

\renewcommand{\bar}{\overline}
\renewcommand{\hat}{\widehat}
\newcommand{\ronum}[1]{\lowercase\expandafter{\romannumeral #1\relax}}

\def\^{{\wedge}}
\def\*{{\star}}

\title{On the Chiral Ring and Vacua of $\CN = 1$ Adjoint SQCD}

\author[\natural]{Ke Ye}

\affiliation[\natural]{Walter Burke Institute for Theoretical Physics, 
California Institute of Technology, \\Pasadena, CA 91125, USA}

\emailAdd{kye@caltech.edu}

\abstract{ We analyze classical and quantum chiral ring relations of four dimensional $\CN = 1$ adjoint SQCD with superpotential turned on for the adjoint field. In particular, for the mass deformed theory  we obtain the complete on shell vacuum expectation value for various gauge invariant chiral operators and find non-trivial gaugino condensations. When approaching to massless limit nontrivial flat directions in the moduli space of vacua appear, where the Coulomb branch can be naturally classified and the Higgs branch receives quantum corrections. We argue that the solution of the chiral ring is in one-to-one correspondence with supersymmetric vacua, provided that an additional Konishi anomaly equation is included.
\\
\\
\\
\\
\\
\\
{\tt CALT-TH-2017-029}
}

\begin{document}

\maketitle

\section{Introduction}

The realm of $\CN = 1$ supersymmetric theories in four dimensions exhibits various interesting phenomena, among which electric magnetic dualities play an important role. The pioneering work of Seiberg \cite{Seiberg:1994pq} demonstrated the IR equivalence of two seemingly distinct gauge theories, in which he showed several exact matchings between operators, moduli space of vacua and massless excitations near singularities. This provides many insights into the non-abelian gauge dynamics of $\CN = 1$ theories. 

Soon it was realized that such dualities are generic for $\CN = 1$ theories \cite{Intriligator:1995id,Intriligator:1995ne}. In \cite{Kutasov:1995np,Kutasov:1995ve}, an attempt was made by Kutasov to analyze the dynamics of $\CN=1$ SQCD with fundamental matter plus one adjoint chiral multiplet (ASQCD)\footnote{In the rest of the paper, we will call the ASQCD with tree level superpotential considered in \cite{Kutasov:1995ve} for adjoint superfield ``Kutasov model".}. He showed that by properly adding a superpotential term for adjoint chiral multiplet that truncates the chiral ring of the theory, a generalized version of Seiberg duality also exists. This duality undergoes various semi-classical consistency checks \cite{Kutasov:1995ss}, and it also sheds light on the quantum chiral ring relations in the original electric theory: a quantum chiral ring relation for Coulomb operators are in fact classical combinatoric constraints in the dual theory. The duality was further explored by \cite{Csaki:1998fm,Murayama:2001ch} to understand the spectra of the confining theory; the corresponding effective superpotential were written down. It was shown there that the effective superpotential is generated by multi-instanton effects in the dual theory.

Meanwhile, another important progress was achieved by the seminal work of Dijkgraaf and Vafa \cite{Dijkgraaf:2002dh} in probing $\CN=1$ dynamics. They conjectured that the effective superpotential of a wide class of ${\CN = 1}$ supersymmetric gauge theories can actually be calculated perturbatively in a closely related matrix model, whose potential is just the classical superpotential of the gauge theory. A striking conclusion was that only \textit{planar} diagrams in the matrix model suffice. Later, Cachazo $et\ al$ \cite{Cachazo:2002ry} provided a purely field-theoretic proof of the correspondence proposed by Dijkgraaf and Vafa, by analyzing Konishi anomalies and chiral rings of $U(N)$ gauge theory with one adjoint chiral multiplet. The powerful conjecture of \cite{Dijkgraaf:2002dh} makes many exact computation in $\CN = 1$ theories (with or without adjoint superfield) accessible; to name several but not all of them, see for instance, \cite{Seiberg:2002jq, Cachazo:2003yc, McGreevy:2002yg, Berenstein:2002sn, Argurio:2002xv, Suzuki:2002gp, Feng:2002zb, Feng:2002yf, Feng:2002is, Klein:2003we}.

However, even with the proposal of duality and the tools from matrix model, there are many other peculiar phenomena in ASQCD that escape precise understanding. For instance, with the aid of $a$-maximization \cite{Intriligator:2003jj, Barnes:2004jj, Kutasov:2003ux}, one discovers that for Kutasov model at large $N$, some chiral operators decouple and become free under RG flow, introducing in the IR so-called ``accidential symmetry"\cite{Kutasov:2003iy}. Moreover, in \cite{Gukov:2015qea} the author found that in such class of theories there are UV irrelevant operators whose scaling dimensions cross marginality under the flow, hence are ``dangerously irrelevant" \cite{Gukov:2016tnp}. The appearance of such operators are quite counter-intuitive in the sense that in the Morse theory interpretation, RG flow is usually triggered by relevant operators; in other words, the relevant operators are ``consumed" along the RG trajectory, and its number should thus decrease along the flow. This ``marginality crossing" behavior is in fact special only to $\CN = 1$ theories in four dimensions; indeed, as shown in \cite{Argyres:2015ffa}, $\CN = 2$ theories do not admit dangerously irrelevant operators. 

Resolving these peculiarities in $\CN = 1$ ASQCD often requires a more precise understanding of vacuum structure, and it is our main motivation of this paper. We will focus on chiral rings of Kutasov model as well as its mass deformed counterpart. The chiral ring probes the vacua of the theory, and tells us about the quantitative behavior at low energies: $e.g.$, chiral symmetry breaking, confinement and electric-magnetic duality. The complete chiral ring relation for $U(N)$ theory with one adjoint chiral superfield is obtained in \cite{Svrcek:2003az}, and our work is a generalization of that. 

We remark that Kutasov model falls into an ADE classification of SQCD with adjoints \cite{Intriligator:2003mi}. This series are revisited recently in \cite{Intriligator:2016sgx}, where some puzzles are found. We hope that the full analysis of quantum chiral ring would resolve these puzzles and, eventually helps understand the entire ADE series\footnote{See for instance, \cite{Mazzucato:2005fe} on some related work.} or ASQCD without superpotentials.

\subsection{Background and summary}

In this paper we analyze the chiral ring of four dimensional $\CN = 1$ supersymmetric $U(N_c)$ gauge theory with one chiral multiplet $\Phi$ in the adjoint representation of $U(N_c)$, and $N_f$ fundamental as well as antifundamental chiral multiplets ${\tilde Q}_{\tilde f}$ and $Q^f$ where $f, \tilde f = 1,2,\dots, N_f$. We consider asymptotic free theories, namely $2N_c > N_f$. The Lagrangian of the theory is
\begin{equation}
\mathcal{L} = \frac{1}{g^2} \bbra{\int d^4 \theta\ Q^{\dagger}_i e^V Q^i + \tilde{Q}_{\tilde i} e^{-V} \tilde{Q}^{\dagger \tilde i} + \Phi^{\dagger} e^{[V, \cdot]} \Phi } + \frac{1}{ 4 g^2}\pbra{ \int d^2 \theta\ W^{\alpha}W_{\alpha} + c.c.},
\end{equation}
where for simplicity we do not distinguish between the $U(1)$ couplings in $U(N_c)$ and $SU(N_c)$ couplings, unlike that of \cite{Shifman:2009mb}. We also think of $U(N_c)$ Kutasov model as coming from $SU(N_c)$ model by gauging the $U(1)$ baryon symmetry. Kutasov model also requires a superpotential of $\Phi$ labelled by a positive integer $k$,
\begin{equation}
W(\Phi) = \frac{h}{k+1} {\rm Tr} {\Phi}^{k+1}.
\label{superpotentialX}
\end{equation}
and the UV theory enjoys an $SU(N_f)_L \times SU(N_f)_R \times U(1)_r$ symmetry. In this paper, we mostly focus on $k=2$.

For $kN_f < N_c$, the theory does not have a quantum vacua; for $kN_f = N_c$ the vacua is modified quantum mechanically; for $kN_f = N_c + 1$ the theory is s-confining, and the effective potential is given by a set of composite degrees of freedom with an irrelevant potential. For $kN_f > N_c$ the theory admits a dual magnetic description with gauge group $U(kN_f - N_c)$.

Kutasov model in general has nontrivial moduli spaces, to understand its quantum chiral ring/quantum vacua, one adds proper deformations to the tree level potential \eqref{superpotentialX} to collapse the flat directions. The most general single trace deformation we can add is \cite{Seiberg:2002jq, Cachazo:2003yc}
\begin{equation}
W_{{\rm tree}} = {\Tr} {\tilde W}(\Phi) + {\tilde Q}_{\tilde f} m^{\tilde f}_f (\Phi) Q^f,
\label{deformation}
\end{equation}
where 
\begin{subequations}
\begin{align}
& {\tilde W}(z) = \sum_{n=0}^k \frac{1}{n+1} g_n z^{n+1} \label{deformPhi},\\[0.5em]
& m_f^{\tilde f} (z) = \sum_{n=1}^{l+1} m^{\tilde f}_{f,n} z^{n-1}.\label{deformQ}
\end{align}
\end{subequations}
Also we define $L = l N_f$.

We will call such theory with deformed superpotential \eqref{deformation} the ``mass deformed" version or ``deformed cousin" of Kutasov model. In the bulk of the paper we will be frequently comparing massive and massless theories. 

The paper is organized as follows. In section \ref{sec2:chiral ring} we review some well-known facts about the chiral ring for $U(N_c)$ ASQCD. We classify chiral operators and describe their relations, with special emphasis on two equivalent descriptions: the algebraic description in terms of generators and relations, as well as the geometric description in terms of expectation values for various composite fields. 

In section \ref{sec3:classical} we calculate the the classical chiral ring and describe different branches of the moduli space.

After that, section \ref{sec4:quantum} is devoted to understand the quantum corrections to the chiral ring. We will list the complete Konishi anomaly equations that give the perturbative chiral ring. The nonperturbative corrections come from certain resolvent operators, whose periods over one cycles of some auxilliary Riemann surface should be integer \cite{Cachazo:2003yc}. It has also been known how to solve the off-shell vacuum expectation values for mass deformed theory \cite{Cachazo:2003yc}; and in this paper we solve them \textit{on-shell}. In the mass deformed theory, the classical vacua are shifted by quantum effects and there are nonvanishing gaugino condensations. With the inclusion of a new Konishi anomaly equation, we are able to prove that the solutions of the chiral ring are in one-to-one correspondence of the supersymmetric vacua. Then, we focus on massless Kutasov model itself. The difficulty of understanding the flat direction of the moduli, unlike that of SQCD, is that the theory has more possible deformations. We will examine a special massless limit and its implications. 

Finally, section \ref{sec5:examples} applies the established framework to some examples of massless model. We will see the existence of quantum corrections directly.

\section{Chiral rings in $\CN = 1$ theories}\label{sec2:chiral ring}

Following the notation of \cite{Cachazo:2002ry,Seiberg:2002jq} we review some basics of chiral rings of four dimensional $\CN = 1$ theories, with fundamental plus adjoint matter. An operator $\CO$ is \textit{chiral} if it is annihilated by a pair of supercharges of the same chirality: $[{\bar Q}_{\dot \alpha}, \CO \} = 0$. One readily checks that a product of two chiral operators is again a chiral operator, therefore chiral operators form a ring.

In the chiral ring, one defines an equivalence relation, namely two chiral operators are equivalent if they differ by a ${\bar Q}_{\dot \alpha}$-exact term. Modulo this equivalence relation, a chiral operator is independent of the position since
\be
\frac{\partial}{\partial x^{\mu}} {\cal O}(x) = \left[ P^{\mu}, {\cal O}(x) \right] = \{ {\bar Q}^{\dot \alpha}, \left[ Q^{\alpha}, {\cal O}(x) \right] \}.
\ee
Therefore, the correlation function of the form $\langle {\cal O}^{(1)} (x_1) {\cal O}^{(2)} (x_2) \dots {\cal O}^{(n)} (x_n) \rangle$ is independent of each coordinate $x_1, x_2, \dots x_n$. It is then possible to move each operator insertion to be mutually far away, such that the expectation value factorizes:
\be
\langle {\cal O}^{(1)} (x_1) {\cal O}^{(2)} (x_2) \dots {\cal O}^{(n)} (x_n) \rangle = \langle {\cal O}^{(1)} \rangle \langle {\cal O}^{(2)} \rangle \dots \langle {\cal O}^{(n)} \rangle.
\ee

For ASQCD, we need to classify all the possible chiral operators modulo ${\bar Q}_{\dot \alpha}$-exact terms. A crucial fact used in \cite{Cachazo:2002ry, Argurio:2003ym} is that, for an adjoint valued chiral superfield ${\cal O}$,
\be
\left[{\bar Q}_{\dot \alpha}, D_{\alpha \dot \alpha} {\cal O} \right \} = [ W_{\alpha}, \CO \},
\ee
which implies the adjoint superfield $\Phi$ commutes with vector superfield $W_{\alpha}$ while $W_{\alpha}$ anti-commutes with $W_{\beta}$. For fundamentals, $W_{\alpha}Q^f$ as well as ${\tilde Q}_{\tilde f} W_{\alpha}$ is not in the chiral ring \cite{Seiberg:2002jq}. Therefore the possible candidates for the ring are
\begin{subequations}
\begin{align}
u_k & =  {\Tr}\Phi^k, \label{Casimir}\\[0.5em]
w_{\alpha,k} & =  \frac{1}{4\pi}  {\Tr}\Phi^k W_{\alpha}, \\[0.5em]
r_k & =  - \frac{1}{32 \pi} {\Tr}\Phi^k W_{\alpha}W^{\alpha},\\[0.5em]
v^f_{{\tilde f},k} & =  {\tilde Q}_{\tilde f} \Phi^k Q^f.
\end{align}
\label{operators}
\end{subequations}
We name $u_k$ the Casimir operators, $r_k$ the generalized glueballs, $w_{\alpha,k}$ the generalized photinos, and $v_k$ the generalized mesons\footnote{There is a slight notation difference between here and what people usually call ``generalized mesons" in the literature. What we mean by $v_k$ is often denoted as $M_{k+1}$.}. Their form suggests to define resolvent operators as the generating function of these chiral operators
\begin{subequations}
\begin{align}
T(z)& =  {\Tr} \frac{1}{z - \Phi},\\[0.5em]
w_{\alpha}(z) & = \frac{1}{4\pi} {\Tr} \frac{W_{\alpha}}{z - \Phi},\\[0.5em]
R(z) & =  -\frac{1}{32\pi^2} {\Tr} \frac{W_{\alpha}W^{\alpha}}{z - \Phi},\\[0.5em]
M_{\tilde f}^{f}(z) & =  {\tilde Q}_{\tilde f} \frac{1}{z - \Phi} Q^f.
\end{align}
\label{resolvent}
\end{subequations}
We will be mostly interested in the resolvent $T(z)$, $R(z)$ and $M(z)$. For supersymmetric vacua, the chiral operators $w_{\alpha,k}$ are zero \cite{Svrcek:2003az}. Although there are nontrivial ring relations among $w_{\alpha, k}$, for solving the vacua we can temporarily neglect them, see section \ref{sec4:quantum}. For $U(N_c)$ theories, the single baryon $B^{[i_1,\dots, i_k][i_{k+1},\dots, i_{N_c}]}$ formed by dressed quark is not gauge invariant; but the composite ${\tilde B}B$ is. However, such operators are not in the chiral ring since they can be expressed in terms of generalized mesons, thus are not independent.

In general, whether at classical or quantum level, the chiral ring of a theory $\CT$ is a quotient of polynomial ring by some ideal, $\CS$:
\be
\CR(\CT) = \mathbb{C} [u_k, w_{\alpha,k}, r_k, v^f_{{\tilde f},k}] / \CS.
\label{general ring}
\ee
We call the ideal $\CS$ the chiral ring relation. Such notation provides two interpretations of the chiral rings. First, the solution satisfying the relation given by $\CS$ parametrize the supersymmetric vacua. Hence one thinks of the moduli space of vacua as an algebraic variety defined by ideal $\CS$ in the polynomial ring. Second, the chiral ring is the coordinate ring defined on the variety. These two interpretations establish an algebraic and geometric connections between chiral rings and vacua of the theory, similar to the stories in classical algebraic geometry.

Specifically, let $\mathsf{V}(\cdot)$ denote the operation of taking algebraic varieties of an ideal, $\mathsf{I}(\cdot)$ the operation of taking polynomials vanishes on the algebraic variety, then by Hilbert's Nullstellensatz, 
\be
\mathsf{I}(\mathsf{V}(\CS)) = \sqrt{\CS},
\label{Nullstellensatz}
\ee
with $\sqrt{\CS}$ the radical ideal. In modern language of schemes, we have $\mathsf{V}(\CS) := \text{Spec}\, \CR$.

A remark is in order. Unlike (twisted) chiral ring in two dimensions, in four dimensions the $\CN = 1$ chiral ring cannot be formulated in term of cohomology \cite{Gukov:2015gmm}. The intuitive reason for that is the supercharges (of the same chirality) as part of the definition in the cochain complex carries Lorentz indices, which are rotated into each other under $SO(4)$ Lorentz group. Since one may construct an example that cohomological description fails for a particular supercharge ${\bar Q}_{\dot 1}$, one sees that it fails for all linear combination of two supercharges $a^{\dot \alpha} {\bar Q}_{\dot \alpha}$.

In what follows, we denote $\hat \CS$ as the quantum relations of Kutasov model, and correspondingly $\hat \CR$ for quantum chiral rings.
 
\section{Classical chiral rings of Kutasov model}\label{sec3:classical}

\subsection{Generalities}\label{subsec:generalities}

In this subsection we mainly focus on the massless model with superpotential \eqref{superpotentialX}. We will briefly comment on its relation to the mass deformed counterpart at the end.

From the Lagrangian of the theory we know the corresponding $D$-term equation reads
\begin{equation}
[\Phi^{\dagger}, \Phi] + (Q^i Q^{\dagger}_ i - {\tilde Q}^{\dagger \tilde i} {\tilde Q}_{\tilde i} ) = 0,
\end{equation}
while the $F$-term constraint is
\begin{equation}
\Phi^k = 0,
\end{equation}
so $\Phi$ is nilpotent\footnote{This is not true for $SU(N_c)$ theories, where a traceless condition should be imposed. This additional constraint makes $\Phi$ either diagonalizable or nilpotent. See \cite{Csaki:1998fm,Murayama:2001ch} for more details.} with degree $k$. The nilpotent matrix always has degree no bigger than its order, so for simplicity we only discuss $k \leq N_c$ in this paper\footnote{Strictly speaking, $k = N_c$ case is in fact a double trace superpotential, as $\Tr X^{N_c+1}$ is not independent.}. The only nilpotent matrix which is diagonalizable is zero matrix; others can only be put into Jordan normal form:

\begin{equation}
\Phi = \left( \begin{array}{cccc} J_1 &  &  & \\ & J_2 &  & \\ & & \ddots & \\ & & & J_n \end{array} \right),
\label{nilpotent}
\end{equation}
where the Jordan block $J_i$ is
\begin{equation}
J_i = \left( \begin{array}{cccc} \lambda_i & 1 & & \\ & \lambda_i & 1 & \\ & & \ddots & 1 \\ & & & \lambda_i \end{array} \right).
\end{equation}
The nilpotency implies that $\lambda_1 = \lambda_2 = \dots =\lambda_n = 0$. A Jordan block $J_i$ is uniquely determined by its order $N_i$. Thus a nilpotent matrix can be labelled by a partition of $N_c$, $[N_1, N_2, \dots, N_n]$, characterizing the size of Jordan blocks : $N_1 + N_2 + \dots + N_n = N_c$ with $k \geq N_1 \geq N_2 \geq \dots \geq N_n$. We use the symbol $Y$ as a Young tableau with $i$-th row of length $N_i$. It is a Young tableau for partition of $N_c$ into integers no larger than $k$.

For nilpotent matrix, we always have
\begin{equation}
{\Tr} \Phi^j = 0, \ \ \ \ j>0.
\label{trivialCasimir}
\end{equation}
which means classically, the vevs of Casimir operators $u_j$ in \eqref{Casimir} are always zero. Note this does not mean $u_j = 0$ \textit{in the chiral ring}\footnote{In mathematical language, the two coordinate ring may define the same classical algebraic varieties, but they do not define the same scheme.}. In the meantime, the vevs of generalized glueballs $r_j$ are in general proportional to the strong coupling scale $\Lambda^{2N_c - N_f}$, and are constrained by fermionic statistics. Since they can only be formulated using adjoint $\Phi$ and vector superfield $W_{\alpha}$ as in \eqref{operators}, the constraints are exactly the same as that in \cite{Svrcek:2003az} and we will not include them in current analysis. Therefore, modulo generalized glueballs and photinos, the classical chiral ring of $U(N_c)$ Kutasov model is a quotient ring of the polynomial ring generated by generalized mesons and Casimir operators:
\be
\CR_{N_c, N_f, k} = {\mathbb{C}} \left[u_1, u_2, \dots, u_{k-1}, v_{0,\tilde f}^f, v_{0,\tilde f}^f, \dots, v_{k-1,\tilde f}^f \right] / {\CS \pbra{u_1, u_2, \dots, u_{k-1}, v_0, v_1, \dots, v_{k-1}}}.
\label{classical ring}
\ee

The constraint $\CS(u_1, u_2, \dots, u_{k-1}, v_0, v_1, \dots, v_{k-1})$ is hard to compute in general. A powerful tool that helps is from computational algebraic geometry. To be more specific, classically we can form a quotient ring using microscopic fields:
\be
{\CR}_{\text{micro}} = \mathbb{C} \left[ {\tilde Q}_{\tilde f}^{\alpha}, Q^f_{\alpha}, \Phi^{\alpha}_{\beta} \right] / {\CS_F},
\ee
where $\CS_F$ comes from $F$-term equations of the superpotential. We do not have to consider the $D$-term once we complexify the gauge group \cite{Luty:1995sd}. The vacuum is parameterized by gauge invariant data, $c.f.$ equation \eqref{general ring}. The natural map arising from composing microscopic field into gauge invariant ones extends to a map between rings:
\be
\psi:  \mathbb{C} \left[u_k, v_{k, \tilde f}^f \right] \rightarrow \CR_{\text{micro}}.
\label{ringMap}
\ee
Then by definition
\be
\CS = \ker \psi.
\ee
Computation of this kernel is standard in the theory of Gr\"{o}bner basis \cite{cox2006using,cox1992ideals}. This method has already been adopted in understanding the vacua and computing Hilbert series of the vacuum moduli, see $e.g.$ \cite{Gray:2008yu,Gray:2006jb}. In section \ref{subsec3.2}, we will explicitly see how this works.

The above algebraic construction is quite abstract. We now turn to concrete description in terms of the moduli space of vacua. As we already know, the Coulomb branch vev $\langle \Phi \rangle$ is parametrized by Young tableau $\left[ N_1, N_2, \dots, N_n \right]$. There are two cases to consider:

\begin{enumerate}

\item[(1)] When all $N_i = 1$. The $D$-term equation becomes that of SQCD with fundamental matter, and there is nontrivial Higgs branch. For $kN_f > N_c+1$, at the root of the Higgs branch the theory is conjectured to be in non-abelian Coulomb phase \cite{Kutasov:1995ss}. 

\item[(2)] $N_i > 1$ for some $i$. Since nontrivial Jordan block does not commute with its conjugate, in general the vevs of quark superfields $\langle Q \rangle$ and $\langle \tilde Q \rangle$ are not zero. We will call it the mixed branch.

\end{enumerate}

In \eqref{trivialCasimir} we see the vevs of gauge invariant Casimir operators are always zero. However the above two cases reveal there are distinct branches in the vacuum moduli. Then the natural question is how can one distinguish between them. Classically, we might tell which branch we are in by looking at the flat directions of generalized mesons. In the branch $[1,1,\dots, 1]$ only $v_0$ is nontrivial, but for other branches more non-trivial generalized mesons appear. However, we will not use such descriptions because such flat directions receive quantum corrections.

\begin{figure*}[htbp]
\begin{adjustwidth}{-0.0cm}{}
        \centering
      \begin{subfigure}[t]{.3\textwidth}
        \includegraphics[width=3cm]{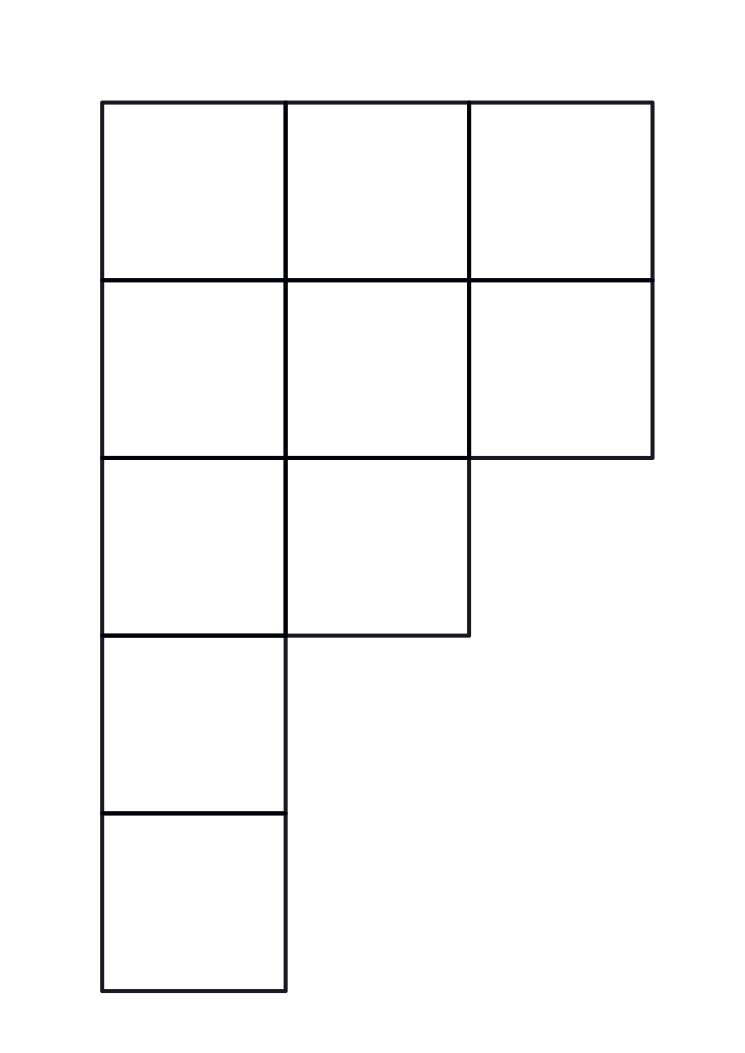}
        \caption*{(a)}
      \end{subfigure}
      \hspace{1.5cm}
      \begin{subfigure}[t]{.3\textwidth}
        \includegraphics[width=4cm]{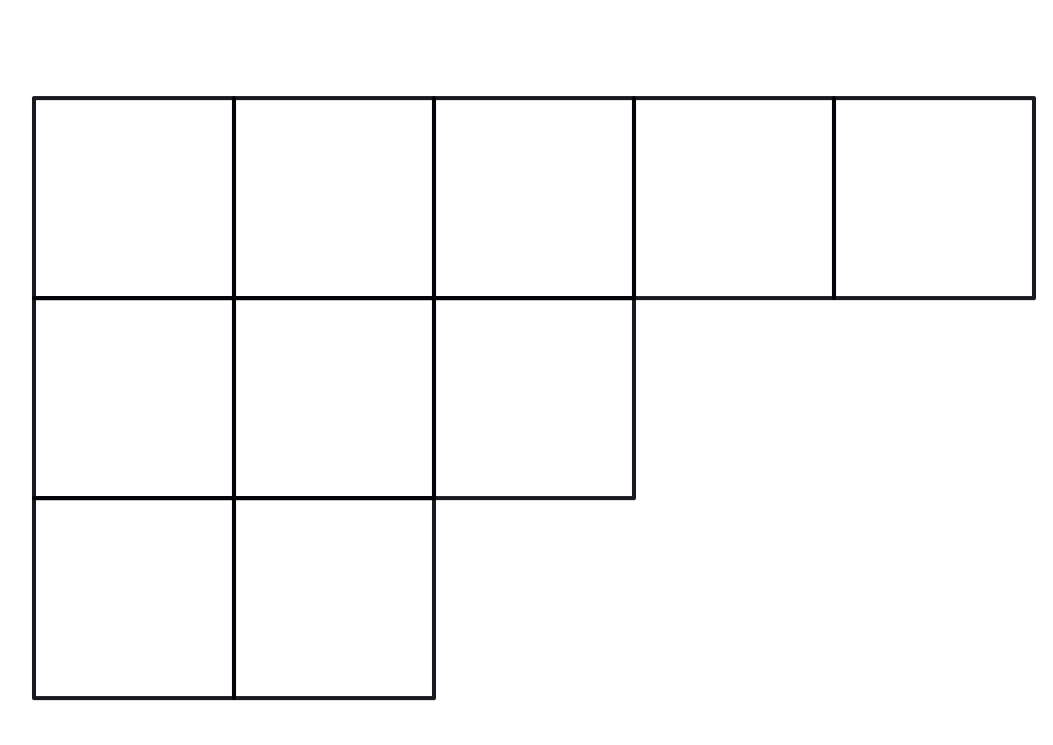}
        \caption*{(b)}
      \end{subfigure}
  \end{adjustwidth}
  \caption{The deformation of nilpotent matrix in the group $U(10)_{\mathbb C} \simeq GL(10)$. In (a) the nilpotent matrix is labelled by $Y = [3,3,2,1,1]$, while the deformed matrix is given by $Y' = [5,3,2]$, with low energy gauge group $U(5) \times U(3) \times U(2)$.}
  \label{YandYD} 
\end{figure*}

Alternatively one can try to study the branch when the deformation \eqref{deformPhi} is turned on. Moreover we require the deformation is sufficiently generic and $g_0 \neq 0$ in \eqref{deformPhi}. It is not hard to see that now $\Phi$ must be diagonalizable, with entries the roots of polynomial
\be
{\tilde W}'(z) = \sum_{n = 0}^k g_n z^n = \prod_{j = 1}^k (z - a_j).
\label{Wderivative}
\ee
Then the Coulomb branch vev $\langle \Phi \rangle$ is labelled by integers $s_1 \geq s_2 \geq \dots \geq s_k$, the number of each root of \eqref{Wderivative}. Therefore we can label this in in terms of another Young diagram $Y'$: $[\dl{s_1}, \dl{s_2}, \dots, \dl{s_k}]$, the partition of $N_c$ into no more than $k$ integers\footnote{We use an underline to remind the reader that they are Young tableau for mass deformed theory.}. It is a standard fact that
\be
Y' = Y^D,
\ee
where $Y^D$ is the dual Young diagram of $Y$. This is also frequently used in the literature as the mapping between nilpotent element and semisimple element. Careful readers now may worry that the mapping is not one-to-one; one can permute the roots $\{a_i\}$ corresponding to the integer $\{s_i\}$. However, there is a natural way to make this mapping one-to-one, due to the fact that their semi-classical unbroken gauge group for a given set of $s_i$ are uniquely fixed regardless of permutation of roots: $U(s_1) \times U(s_2) \times \cdots U(s_k)$. Therefore we may define our map from a nilpotent $\langle \Phi \rangle$ to the image taking the rank of unbroken subgroup of $U(N_c)$. In figure \ref{YandYD} we give an example of the correspondence of the Young diagrams.

This identification is more robust than the previous one in the sense that patterns of unbroken gauge group are rigid against quantum corrections. We will see that it is indeed the case in section \ref{sec4:quantum}.

As we have seen that the deformation \eqref{deformPhi} is important to distinguish between different branches, it is illustrative to summarize what the vacua look like if the full deformation \eqref{deformation} is turned on \cite{Cachazo:2003yc}. In this case, the vacua consist of Coulomb branch (pseudo-confining branch) and Higgs branch. For Coulomb branch, we have
\be
\langle \Phi \rangle = {\rm diag} (a_1, \dots, a_1, a_2, \dots, a_2, \dots, a_k, \dots a_k ), \ \ \ \ \langle {\tilde Q}_{\tilde f} \rangle = \langle Q^f \rangle = 0.
\label{classicalCoulomb}
\ee
For Higgs branch we have
\begin{subequations}
\begin{align}
& \langle \Phi \rangle = {\rm diag} (b, a_1, \dots, a_1, a_2, \dots, a_2, \dots, a_k, \dots a_k ),\\[0.5em]
& \langle {\tilde Q}_{\tilde f}^{\beta} \rangle = \langle Q^f_{\beta} \rangle = 0, \ \ \ \beta = 2,3,\dots N_c,\\[0.5em]
& Q^f_1 \pbra{\sum_{n=1}^{l+1}(n-1) b^{n-2} m^{\tilde f}_{f,n}} {\tilde Q}^1_{\tilde f} + {\tilde W}'(b) = 0,\label{HiggsQ}
\end{align}
\label{classicalHiggs}
\end{subequations}
where $b$ is the root of $B(z) = \det \bbra{m_f^{\tilde f} (z)} = 0$. Similar reasoning to that of \cite{Cachazo:2003yc} reveals that root $b$ can only appear in $\langle \Phi \rangle$ once. The solution can also be elegantly packaged as
\be
M(z) = - \sum_{I = 1}^{l N_f} \frac{r_I {\widetilde W}'(b_I)}{z-b_I} \frac{1}{2\pi i} \oint_{b_I} \frac{1}{m(x)} dx
\label{Mcl}
\ee
where $r_I = 0, 1$ is the number of $b_I$ in the diagonal of $\langle \Phi \rangle$. This solution of classical Higgs branch will be important in section \ref{sec4:quantum}. 

\subsection{Example: $U(2)$ theory with $k = 2$}\label{subsec3.2}

Having discussed generalities, it is time to get refreshed by a couple of examples. In this subsection we will be illustrating the case $N_c = 2$, $k=2$ with $N_f = 1,2$. We have two choices of Young tableau for $\langle \Phi \rangle$: $[1,1]$ or $[2]$. Upon deformations by \eqref{deformPhi}, $[1,1]$ corresponds to the dual vacua $[\dl{2}]$ where the gauge group remains unbroken as $U(2)$, but $[2]$ corresponds to the dual vacua $[\dl{1},\dl{1}]$ where gauge group is broken down to $U(1)^2$. For $[1,1]$ branch, $v_1 = 0$ but it is nonzero for $[2]$. Since $\Phi^2$ vanishes, one concludes that $v_j = 0$ for $j \geq 2$. Therefore we know classically,
\be
{\CR}_{2, N_f, 2} = \mathbb{C} \left[ u_1, v_0, v_1 \right] / \CS (u_1, v_0, v_1).
\ee

Next we turn to the classical relation $\CS$. A nice computer program that produces the kernel of the map $\psi$ in \eqref{ringMap} is \texttt{Macaulay 2} \cite{M2,eisenbud2002computations}. In the following we list the relations $\CS (u_1, v_0, v_1)$ for $N_f = 1, 2$:

\vspace{5pt}
$\bullet\ N_f = 1$. 
\be
{\CR}_{2,1,2} = \mathbb{C} \left[u_1, v_0, v_1 \right] / \langle u_1^3, u_1^2 v_1, u_1 v_1^2, u_1^2 v_0 - 2 u_1 v_1 \rangle.
\ee
Notice that $u_1$ is nilpotent in the chiral ring; the classical relation implies that $u_1 = 0$ as an algebraic variety, and the rest constraints in the relations are trivially satisfied. So $v_0$, $v_1$ take arbitrary complex values.

\vspace{5pt}
$\bullet\ N_f = 2$. It turns out that the relation can be compactly cast as
\be
{\CS}_{2,2,2} = \langle & u_1^3, u_1^2 v_1, v_1 \det v_1,  u_1 \det v_0 - \det (v_0+v_1) + \det v_0 + \det v_1, \\[0.5em] 
& u_1 v_{1,i}^j v_{1,k}^l, u_1^2 v_{0} - 2 u_1 v_{1}, u_1 (v_{0,i}^j v_{1,k}^l - v_{0,k}^l v_{1,i}^j), v_1 \det(v_0 + v_1) - v_1\det v_0 - v_0 \det v_1 \rangle.
\label{R222}
\ee
In solving the chiral ring, we see again that the nilpotent element $u_1 = 0$. What remains are $\det v_1 = 0$, following from the fact $\langle \Phi \rangle$ has rank $1$, and $\det(v_0 + v_1) - \det v_0 = 0$.


\subsection{Examples: $U(3)$ theory with $k = 2$}\label{subsec:classicalU(3)}

Our next example is $U(3)$ theory with $k=2$. Here we only analyze $N_f = 1$. For large numbers of flavors, the relations quickly become very complicated. The adjoint chiral multiplet has two choices of vevs:
\be
\langle \Phi \rangle_{[1,1,1]} = \left( \begin{array}{ccc} 0 & 0 & 0 \\ 0 & 0 & 0 \\ 0 & 0 & 0 \end{array} \right), \ \ \ \ \ \langle \Phi \rangle_{[2,1]} = \left( \begin{array}{ccc} 0 & 1 & 0 \\ 0 & 0 & 0 \\ 0 & 0 & 0 \end{array} \right).
\ee

For $N_f=1$, we again see the chiral ring is generated by $u_1$, $v_0$ and $v_1$ as:
\be
{\CR}_{3,1,2} = \mathbb{C} \left[ u_1, v_0, v_1 \right] / \langle u_1^4, u_1^3 v_1, u_1^2 v_1^2, u_1^3 v_0 - 3 u_1^2 v_1 \rangle.
\ee
The Casimir operator $u_1$ is again nilpotent in the chiral ring.

\subsection{General $U(N_c)$ theory with $k=2$}\label{classicalU(Nc)}

Motivated by our study of $U(2)$ and $U(3)$ theories with $k=2$, we conjecture the classical constraints for general $U(N_c)$ theory with $N_f$ fundamental flavors, with $k=2$ as follows. The superpotential \eqref{superpotentialX} forces the nilpotent matrix $\langle \Phi \rangle$ to be
\be
Y_{N_c, N_f, 2} = [2,2,\dots, 2, 1,1,\dots 1]
\ee
where we denote $n_2$ as number of order $2$ Jordan block, then the trivial Jordan block has number $N_c - 2 n_2$. Apparently, the chiral ring relation should not depend on the choice of $n_2$. For $N_f = 1$, we can write down the complete relations $\CS$; but for other number of flavors, we only write down relations in $\sqrt{\CS}$. They may not necessarily be the true chiral ring relation - as the chiral ring contains nilpotent elements.

\begin{enumerate}
\item[$\bullet$] $N_f = 1$:
\be
\CR_{N_c, 1, 2} = \mathbb{C} [u_1, v_0, v_1] / \langle u_1^{N_c + 1}, u_1^{N_c} v_1,  u_1^{N_c-1} v_1^2,  u_1^{N_c} v_0 - N_c u_1^{N_c -1} v_1 \rangle.
\ee
\end{enumerate}

\begin{enumerate}
\item[$\bullet$] $N_f \leq \lfloor N_c / 2 \rfloor$. The solutions to $\CS$ do not constrain $v_0$ and $v_1$, they can take arbitrary complex values. This may be confirmed in the $N_f = 1$ case above when taking $u_1 = 0$. We thus have $\sqrt{\CS} = \langle u_1 \rangle$.

\item[$\bullet$] $\lfloor N_c / 2 \rfloor < N_f < N_c$. Since the adjoint chiral superfield is built from rank $\leq \lfloor N_c / 2 \rfloor$ data, the second generalized meson becomes degenerate. The solution for $v_1$ satisfies:
\be
v_{1,j_1}^{[ i_1} v_{1,j_2}^{i_2} \cdots v_{1,j_{\lfloor N_c / 2 \rfloor + 1}}^{i_{\lfloor N_c / 2 \rfloor + 1}]} = 0,
\label{NfsNc}
\ee
and there are no additional constraints on $v_0$.

\item[$\bullet$] $N_c \leq N_f$. We define $\tilde v = v_0 + v_1$. In addition to \eqref{NfsNc}, we have
\be
{\tilde v}_{j_1}^{[ i_1} {\tilde v}_{j_2}^{i_2} \cdots {\tilde v}_{j_{N_c}}^{i_{N_c} ]} - {v}_{0, j_1}^{[ i_1} {v}_{0,j_2}^{i_2} \cdots {v}_{0, j_{N_c}}^{i_{N_c} ]} = 0.
\ee
When $N_c < N_f$ we have yet another relation coming from the degeneration of first generalized meson $v_0$:
\be
{v}_{0, j_1}^{[ i_1} {v}_{0, j_2}^{i_2} \cdots {v}_{0, j_{N_c+1}}^{i_{N_c+1}]} = 0.
\ee
\end{enumerate}


\section{Quantum chiral rings}\label{sec4:quantum}

In this section we analyze quantum chiral rings. When dealing with the quantum vacua with nontrivial flat directions, the usual way is to deform the theory, endowing all the matter with a mass and then taking appropriate limit \cite{Seiberg:1994bz}. We thus introduce the deformation \eqref{deformation} first and study the resulting vacuum expectation values of gauge invariant chiral operators; by taking the limit one ends up with some particular point on the vacuum moduli.

We emphasize that such a way recovers vacua as an algebraic variety (or the radical ideal), but not the true chiral ring, by Hilbert's Nullstellensatz \eqref{Nullstellensatz}.

\subsection{Perturbative corrections}\label{subsec:perturb}

The $F$-term constraint from the superpotential is obtained via chiral rotations $X \rightarrow X + \delta X$ where $X$ is some chiral superfield in the Lagrangian. It can also be viewed as conservation law of the current
\be
J = {\rm Tr} {\bar X} e^V \delta X
\ee
with a source term, which is subjected to Konishi anomaly \cite{Konishi:1983hf,Konishi:1985tu} and its generalized versions \cite{Cachazo:2002ry,Alday:2003gb}. If we pick $\delta X = f(\tilde Q, Q, \Phi, W_{\alpha})$ where $f$ is a holomorphic function of its arguments, the conservation equation can be written as
\begin{equation}
{\bar D}^2 J = {\rm Tr} f(\tilde Q, Q, \Phi, W_{\alpha}) \frac{\partial W_{\text{tree}}}{\partial X} +\ \text{anomaly}\ + {\bar D}. (\dots)
\label{anomalous current}
\end{equation}
Here we may drop the $\bar D(\dots)$ term and set to zero the left hand side of \eqref{anomalous current} since it is a ${\bar Q}_{\dot \alpha}$ commutator, therefore zero in the chiral ring. In the Dijkgraaf-Vafa conjecture, the Konishi anomaly equations are identified as the loop equations of the matrix model \cite{Migdal:1984gj}. 

The one-loop anomaly can be computed as that in \cite{Cachazo:2002ry}. For instance, given an adjoint superfield $X$ and its variation as above, we have
\be
\text{anomaly}~ = \sum_{ijkl} A_{ij, kl} \frac{\partial f(\tilde Q, Q, \Phi, W_{\alpha})_{ji}}{\partial \Phi_{kl}},
\ee
where the coefficient $A_{ij,kl}$ is
\be
A_{ij,kl} = \frac{1}{32 \pi^2} \left[ (W_{\alpha}W^{\alpha})_{il} \delta_{jk} + (W_{\alpha}W^{\alpha})_{jk} \delta_{il} - 2 (W_{\alpha})_{il} (W^{\alpha})_{jk} \right].
\ee

For the mass-deformed ASQCD, the five independent Konishi anomaly equations are \cite{Seiberg:2002jq, Cachazo:2003yc}:
\begin{subequations}
\begin{align}
{\Tr} \frac{{\widetilde W}'(\Phi)}{z- \Phi} + {\tilde Q}_{\tilde f} \frac{{m'}^{\tilde f}_f (\Phi)}{z - \Phi} Q^f & = 2 R(z) T(z) + w_{\alpha}(z) w^{\alpha}(z),\label{anomalyC} \\[0.5em]
\frac{1}{4\pi} {\Tr} \frac{{\widetilde W}'(\Phi) W_{\alpha}}{z - \Phi} & = 2R(z) w_{\alpha} (z),\label{anomalyw}\\[0.5em]
-\frac{1}{32\pi^2} {\Tr} \frac{{\widetilde W}'(\Phi) W_{\alpha} W^{\alpha}}{z - \Phi} & = R(z)^2,\label{anomalyr}\\[0.5em]
\lambda_{f'}^f {\tilde Q}_{\tilde f} \frac{m_f^{\tilde f} (\Phi)}{z - \Phi} Q^{f'} & = \lambda_f^f R(z),\label{anomalyM}\\[0.5em]
{\tilde \lambda}^{{\tilde f}'}_{\tilde f} {\tilde Q}_{{\tilde f}'} \frac{m_f^{\tilde f} (\Phi)}{z - \Phi} Q^f & = {\tilde \lambda}_{\tilde f}^{\tilde f} R(z).\label{anomalytM}
\end{align}
\label{anomaly eqs}
\end{subequations}
The right hand side of equation \eqref{anomalyC} - \eqref{anomalytM} is the anomaly at one loop; Setting them to zero reduces to classical $F$-term equations. Expanding both sides of \eqref{anomaly eqs} around $z \rightarrow +\infty$, and comparing coefficients with the same power of $z$ give perturbative corrections to the chiral ring of the massive theory.

There is one more Konishi anomaly. For an arbitrary matrix $h^{\tilde g}_{g}$, we take our chiral rotation to be
\be
\delta \Phi = \frac{1}{z-\Phi} {\tilde Q}_{\tilde g} h^{\tilde g}_{g} Q^g \frac{1}{z-\Phi},
\ee
then we can write down the sixth Konishi anomaly equation:
\be
 {\tilde Q}_{\tilde g} \frac{W'(\Phi)}{(z-\Phi)^2} Q^g + \sum_{n=1}^{l+1} \sum_{m=0}^{n-2} {\tilde Q}_{\tilde f} \frac{\Phi^m}{z-\Phi} Q^g m^{\tilde f}_{f,n} {\tilde Q}_{\tilde g} \frac{\Phi^{n-2-m}}{z-\Phi} Q^f
 = 2 R(z) {\tilde Q}_{\tilde g} \frac{1}{(z-\Phi)^2} Q^g,
\label{6Anomaly}
\ee
where we have removed $h^{\tilde g}_g$ on both sides. We have also dropped terms that contain $W_{\alpha}Q^f$ or ${\tilde Q}_{\tilde f} W_{\alpha}$ since they are not in the chiral ring.

The off-shell quantum Coulomb branch vacua have been solved by Cachazo, Seiberg and Witten as \cite{Cachazo:2003yc} using the anomaly equations \eqref{anomaly eqs}: 
\begin{subequations}
\begin{align}
2R(z) & = {\widetilde W}'(z) - \sqrt{{\widetilde W}'(z)^2 + f(z)}, \label{quantumR}\\[0.5em]
M(z) & = -\sum_{i=1}^n \frac{1}{2\pi i} \oint_{A_i} \frac{R(x)}{x-z} \frac{1}{m(x)} dx, \label{quantumM} \\[0.5em]
T(z) & = \frac{B'(z)}{2B(z)} - \sum_{I=1}^L \frac{y(q_I)}{2y(z)(z-z_I)} + \frac{g(z)}{y(z)}, \label{CSW T}
\end{align}
\label{CSW solution}
\end{subequations}
where $m(x)$ is the abbreviation for $m_f^{\tilde f}(z)$ in \eqref{deformQ} and $y(z)^2 = {\widetilde W}'(z)^2 + f(z)$. Because of $y(z)$, the solution is defined on a genus $k-1$ Riemann surface $\Sigma$. $A_i$ are the cycles that surrounds the $i$-th cut, smearing of the classical Coulomb branch singularity, and $q_I$'s are the point corresponding to Higgs branch in the first sheet of $\Sigma$ as a double cover of complex plane. Finally,
\begin{subequations}
\begin{align}
& f(z) = \frac{1}{8\pi^2} {\rm Tr} \frac{({\widetilde W}'(z) - {\widetilde W}'(\Phi))W_{\alpha}W^{\alpha}}{z - \Phi}\\[0.5em]
& g(z) =\left< {\rm Tr} \frac{{\widetilde W}'(z) - {\widetilde W}'(\Phi)}{z - \Phi} \right > - \frac{1}{2} \sum_{I=1}^L \frac{{\widetilde W}'(z)-{\widetilde W}'(z_I)}{z-z_I}
\end{align}
\end{subequations}
In solving these equations, it is required that when $z$ approaches to $q_I$, the residue of $T(z)$ should be at most one \cite{Cachazo:2003yc}. We conjecture that this condition is encoded in the sixth anomaly equation \eqref{6Anomaly}, which will be clear in section \ref{subsec:physical} and \ref{subsec:onetoone}. Note the above solutions are off-shell, with $f(z)$ being some generic degree $k-1$ polynomial. We will solve these equation on-shell later.

\subsubsection{Exactness of Konishi anomaly}

A natural question to ask is if the Konishi anomaly receives further quantum corrections. Consider first the perturbative higher loop  corrections. The UV coupling $\tau_{UV}$ is replaced by dynamical scale $\Lambda$. To use holomorphy we write down the symmetry when all the couplings as well as scale $\Lambda$ are treated as background superfields. Following \cite{Cachazo:2002ry}, the combination $U(1)_{\theta} = -2 U(1)_{\Phi} /3 + U(1)_R$ is defined for convenience.

\setlength\extrarowheight{6pt}
\begin{table}[htbp]
\centering
\begin{tabular}{| c | c | c | c | c | c | c | c |}
\hline
 & $ \Delta $ & $SU(N_f)_L$ & $SU(N_f)_R$ & $U(1)_A$ & $U(1)_R$ & $U(1)_{\Phi}$ & $U(1)_{\theta}$ \\[0.5em] \hline
$\Phi$ & $1$ &    {\bf 1}         &      {\bf 1}       &     $0$     &      $\frac{2}{3}$  & $1$ & $0$   \\[0.5em] \hline
 $Q$ & $1$ &   $\square$         &     {\bf 1}        &   $1$       &    $\frac{2}{3}$ & $0$  &  $\frac{2}{3}$ \\[0.5em] \hline
$\tilde Q$ &  $1$ &  {\bf 1}       &         ${\overline \square}$       &   $1$        &      $\frac{2}{3}$   & $0$ & $\frac{2}{3}$ \\[0.5em] \hline
 $g_n$ & $2-n$ & {\bf 1} & {\bf 1}  & $0$ &  $\frac{2}{3}(2-n)$ & $-n-1$ & $2$ \\[0.5em] \hline
$m^{\tilde f}_{f,n} $ & $2-n$ & ${\overline \square}$ & $\square$ & $-2$ & $\frac{2}{3}(2-n)$ & $1-n$ & $\frac{2}{3}$ \\[0.5em] \hline
$ W_{\alpha} $ & $\frac{3}{2}$ & {\bf 1} & {\bf 1}& $0$ & $1$ & $0$ & $1$ \\[0.5em]\hline
$\Lambda^{2N-N_f}$ & $2N - N_f$ & {\bf 1} & {\bf 1} & $2N_f$ & $\frac{2}{3}(2N - N_f)$ & $2N$ & $-\frac{2}{3} N_f$ \\[0.5em]\hline
\end{tabular}
\caption{Summary of charge assignments for operators and couplings. Note these charges are chosen so that there are no quantum anomalies.}
\label{chargeTab}
\end{table} 

Consider first $f = \delta \Phi \propto \Phi$. This variation is considered in \cite{Cachazo:2002ry} and is the coefficient of $z^{-2}$ in the expansion of \eqref{anomalyC}. The difference between our case and \cite{Cachazo:2002ry} is we need to worry about the appearance of $m^{\tilde f}_{f,n}$. The right hand side in the expansion \eqref{anomalyC} has terms proportional to $W_{\alpha}^2$, so it is charged $(0,0,2)$ under $U(1)_A \times U(1)_{\Phi} \times U(1)_{\theta}$. Acceptable corrections should not depend on the negative power of couplings since they should vanish if couplings are zero. The only possible terms are $g_n \Phi^{n+1}$, $W_{\alpha}^2$ and $m^{\tilde f}_{f,n}{\tilde Q}_{\tilde f} \Phi^{n-1} Q^f$, but they are already present in one loop. 

The general case when $\delta \Phi \propto \Phi^{m}$ is similar, where the charge under $U(1)_A \times U(1)_{\Phi} \times U(1)_{\theta}$ becomes $(0, m-1, 2)$. The terms already presented in the one-loop expression are $g_n \Phi^{n+m}$, $m^{\tilde f}_{f,n}{\tilde Q}_{\tilde f} \Phi^{n+m-2} Q^f$ and $\sum_{l=0}^{m-1} {\rm Tr} W_{\alpha}^2 \Phi^{m-l-1} {\rm Tr} \Phi^l$, all of which have the right charge.

Likewise we can consider $\delta Q^f \propto \Phi^m Q^f$ which is the $z^{-m-1}$ coefficient in the expansion of \eqref{anomalyM}. As a result similar to previous argument, we see no higher loop correction is possible which is in accordance with symmetry and holomorphy.

Finally, we can consider $\delta \Phi \propto \Phi^m Q^g h_g^{\tilde g} {\tilde Q}_{\tilde g}\Phi^n$ in \eqref{6Anomaly}. For simplicity we illustrate $m = n = 0$ only. This is the $z^{-2}$ coefficient in the expansion. It is charged $(2, -1, 10/3)$ under $U(1)_A \times U(1)_{\Phi} \times U(1)_{\theta}$. Once again, the allowable term ${\tilde Q}_{\tilde f} \Phi^m Q^g m^{\tilde f}_{f,n} {\tilde Q}_{\tilde g} \Phi^{n-2-m} Q^f$ is already there at one-loop.

Nonperturbatively we should study the algebra of chiral rotations and the Wess-Zumino consistency condition on the anomaly \cite{Wess:1971yu}, following the line of \cite{Svrcek:2003kr}. We define the generators of the algebra as
\begin{subequations}
\begin{align}
L_n & = \Phi^{n+1} \frac{\delta}{\delta \Phi}, \\[0.5em]
Q_{n, \alpha} & = \frac{1}{4\pi} W_{\alpha} \Phi^{n+1} \frac{\delta}{\delta \Phi}, \\[0.5em]
R_n & = -\frac{1}{32\pi^2} W_{\alpha}W^{\alpha} \Phi^{n+1} \frac{\delta}{\delta \Phi}, \\[0.5em]
M^f_{f', n} & = \Phi^n Q^f \frac{\delta}{\delta Q^{f'}},\\[0.5em]
{\tilde M}^{{\tilde f}'}_{{\tilde f},n} & = {\tilde Q}_{\tilde f} \Phi^n \frac{\delta}{\delta {\tilde Q}_{{\tilde f}'}}.
\end{align}
\end{subequations}

Classically they satisfy commutation relations which are an extension of Virasoro algebra:
\begin{equation}
\begin{aligned}
\left[ L_m, L_n\right] & = (n-m) L_{m+n},\ \ \ \  [L_m, Q_{n,\alpha}] = (n-m)Q_{n+m,\alpha},\\[0.5em]
[L_m, R_n] & = (n-m) R_{m+n}, \ \ \ \ \{Q_{m,\alpha}, Q_{n,\alpha} \} = -\epsilon_{\alpha \beta} (n-m) R_{n+m}, \\[0.5em]
[Q_{m,\alpha}, R_n] & = 0 , \ \ \ \ \ \ \ \ \ \ \ \ \ \ \ \ \ \ \ \ \ \ \ \ \ \ [R_m, R_n] = 0, \\[0.5em]
[M^f_{f', n}, M^g_{g', m}] & = \delta^g_{f'} M^f_{g', n+m} - \delta^f_{g'} M^g_{f', n+m}, \\[0.5em]
[{\tilde M}^{{\tilde f}'}_{{\tilde f},n},{\tilde M}^{{\tilde g}'}_{{\tilde g},m}] & = \delta^{{\tilde f}'}_{\tilde g}{\tilde M}^{{\tilde g}'}_{{\tilde f},n+m} - \delta^{{\tilde g}'}_{\tilde f}{\tilde M}^{{\tilde f}'}_{{\tilde g},n+m}, \\[0.5em]
[M^f_{f', n}, {\tilde M}^{{\tilde g}'}_{{\tilde g},m}] & = 0, \\[0.5em]
[L_n, M^f_{f', m}] & = m M^f_{f', n+m}, \ \ \ \ \ \ \ \ \ \ [L_n, {\tilde M}^{{\tilde f}'}_{{\tilde f},m}] = m {\tilde M}^{{\tilde f}'}_{{\tilde f},m+n},\\[0.5em]
[Q_{n, \alpha}, M^f_{f', m}] & = 0, \ \ \ \ \ \ \ \ \ \ \ \ \ \ \ \ \ \ \ \ \ \ \ \ \ [Q_{n, \alpha}, {\tilde M}^{{\tilde f}'}_{{\tilde f},m}] = 0,\\[0.5em]
[R_{n}, M^f_{f', m}] & = 0, \ \ \ \ \ \ \ \ \ \ \ \ \ \ \ \ \ \ \ \ \ \ \ \ \ [R_{n}, {\tilde M}^{{\tilde f}'}_{{\tilde f},m}] = 0.\\[0.5em]
\end{aligned}
\end{equation}
One can in principle include the generator
\be
K_{s, t} & = \Phi^s {\tilde Q}_{\tilde g} h^{\tilde g}_g Q^g \Phi^t \frac{\delta}{\delta \Phi};
\ee
here we do not consider the algebra involving $K_{s,t}$ since when acting on generalized mesons the transformation is not linear anymore. Note due to the presence of fundamentals, there is no $U(1)$ shift symmetry, unlike the case with adjoint only. In terms of these operators, the Konishi anomaly can be expressed as a representation of the algebra:

\begin{equation}
\begin{aligned}
& L_n W_{\text{eff}} = {\cal L}_n, \ \ \ \ \ \ \ \ \ \ \ \ M^f_{f', n} W_{\text{eff}} = {\cal M}^f_{f', n},\\[0.5em]
& Q_{n,\alpha} W_{\text{eff}} = {\cal Q}_{n, \alpha}, \ \ \ \ \ \ \ {\tilde M}^{{\tilde f}'}_{{\tilde f},n} W_{\text{eff}} = {\cal {\tilde M}}^{{\tilde f}'}_{{\tilde f},n},\\[0.5em]
& R_n W_{\text{eff}} = {\cal R}_n.
\end{aligned}
\end{equation}

It is not hard to check that these perturbative anomalies ${\cal L}$, ${\cal Q}$, ${\cal R}$, ${\cal M}$ and ${\cal {\tilde M}}$ satisfy the Wess-Zumino consistency conditions and thus form a representation of the chiral rotation algebra. 

Now we are ready to check the nonperturbative corrections both to the algebra and the Konishi anomalies. Our theory has an axial $U(1)_A$ symmetry. The generators $L$, $Q$, $R$, $M$ and ${\tilde M}$ all have charge $0$ under the $U(1)_A$. Then the correction to the commutation relations should not carry $U(1)_A$ charge as well. But the scale $\Lambda^{2N - N_f}$ has charge $2N_f$. The only way to cancel it is to use powers of $m^{\tilde f}_{f,k}$. To extract singlet from the flavor symmetry, we have to antisymmetrize the indices:
\begin{equation}
\epsilon_{\, {\tilde i}_1 {\tilde i}_2 \dots {\tilde i}_{N_f}} \epsilon^{i_1 i_2 \dots i_{N_f}} m^{{\tilde i}_1}_{i_1, n_1} m^{{\tilde i}_2}_{i_2, n_2} \dots m^{{\tilde i}_{N_f}}_{i_{N_f}, n_{N_f}}.
\end{equation}
When those $m$'s are finite, one expects that all the non-perturbative corrections can be absorbed into redefinition of the elements in the algebra \cite{Ferrari:2007wa, Ferrari:2007bz}. We leave the detailed proof to the future work.

\subsection{Nonperturbative corrections}\label{subsec:nonperturb}

There are other relations in the chiral ring of nonperturbative origin, and typically involving strong coupling scale. Recall that our gauge group is of finite rank, the Casimir operators $\{u_i = {\rm Tr}\Phi^i \}_{i=0}^{+\infty}$ are not all independent. The constraint comes from the characteristic polynomial of matrix $\Phi$:
\be
u_{N_c+p} =  {\cal F}(u_1, u_2, \dots, u_{N_c-1}, u_{N_c}), \ \ \ \ p \in \mathbb{Z}^+.
\label{CasN+p}
\ee
Classically, if we denote $P(z) = \det (z - \Phi) = z^{N_c} + p_1 z^{N_c-1} + \dots +p_{N_c - 1}z + p_{N_c}$ as the characteristic polynomial, then the above relation can be packaged as
\be
\frac{P'(z)}{P(z)} = T(z).
\label{clCasrelation}
\ee
The left hand side of \eqref{clCasrelation} depends on finite number of parameters $p_1, \dots, p_{N_c}$ while the right hand side of \eqref{clCasrelation} contains all the Casimir operators. This implies the classical constraint \eqref{CasN+p}.

\begin{figure*}[htbp]
\begin{adjustwidth}{-2.0cm}{}
        \centering
      \begin{subfigure}[t]{.4\textwidth}
        \includegraphics[width=8cm]{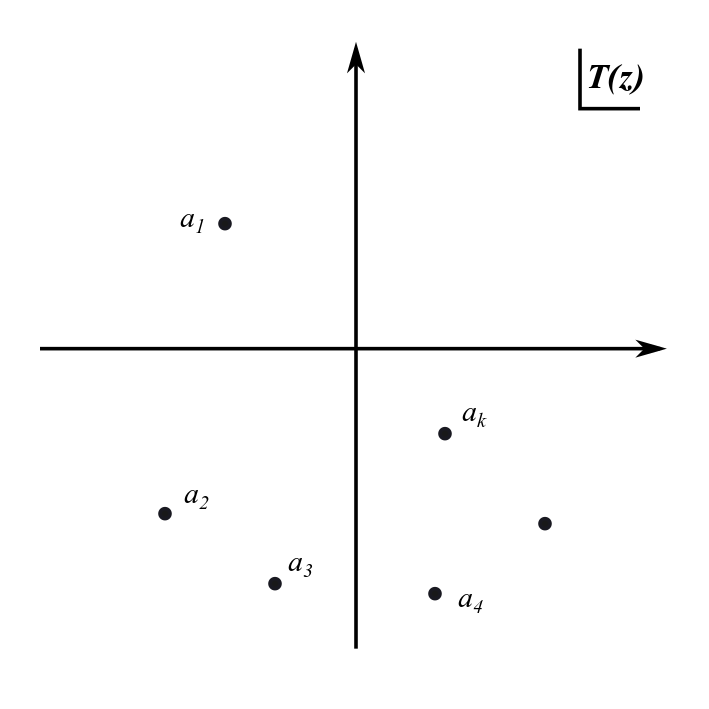}
        \caption*{(a)}
      \end{subfigure}
      \hspace{1.5cm}
      \begin{subfigure}[t]{.4\textwidth}
        \includegraphics[width=8cm]{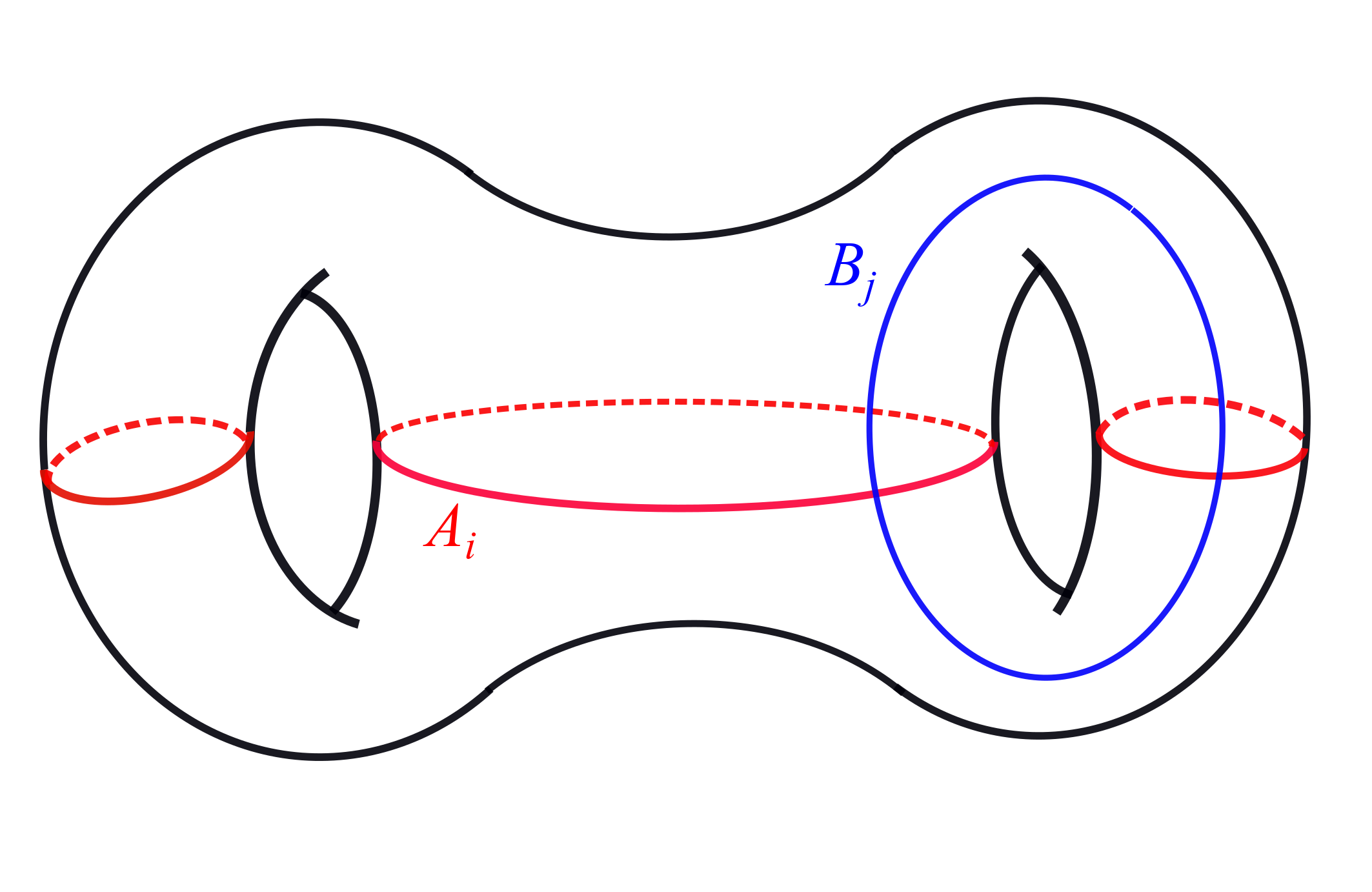}
        \caption*{(b)}
      \end{subfigure}
  \end{adjustwidth}
  \caption{The classical (a) and quantum (b) picture of describing resolvent $T(z)$. Classically, $T(z)$ takes value on a complex plane, with poles located at the root of \eqref{Wderivative}. Quantum mechanically, the complex plane becomes a Riemann surface described by $y(z)^2 = {\widetilde W}'(z)^2 + f(z)$; the poles $a_i$ becomes cuts $A_i$. We also choose $B_i$ that intersects only $A_i$. The quantization condition is around the cycle $A_i$ and $B_j$.}
  \label{TzPlane} 
\end{figure*}

Quantum mechanically \eqref{CasN+p} gets modified by instanton effects, turning into
\be
u_{N_c+p} =  {\hat {\CF}}(u_1, u_2, \dots, u_{N_c-1}, u_{N_c}; \Lambda^{2N_c - N_f}), \ \ \ \ p \in \mathbb{Z}^+.
\label{QCasN+p}
\ee
This can be deduced based on the fact that the resolvent $T(z)$ has quantized periods. Indeed, if we focus on the classical Coulomb branch solution \eqref{classicalCoulomb}, then $T(z)$ has a pole when $z$ approaches to one of the root $a_i$ with residue equal to number of entries of $a_i$. Integrate around small cycle around $a_i$ we have
\begin{equation}
\frac{1}{2\pi i} \oint_{a_i} T(z) dz = N_i  \in \mathbb{Z}.
\end{equation}
Quantum mechanically the poles $a_i$ are smeared into cuts $A_i$, and the complex plane becomes a Riemann surfaces $\Sigma: y(z)^2 = W'(z)^2+f(z)$ \eqref{CSW solution}, but the quantization condition is the same \cite{Cachazo:2003yc}:
\begin{equation}
\frac{1}{2\pi i} \oint_{A_i} T(z) dz = N_i  \in \mathbb{Z},
\label{quantizeA}
\end{equation}
still giving the rank of unbroken gauge group. Hence the rank is robust against quantum corrections, in accordance with what we mentioned in section \ref{subsec:generalities}. See figure \ref{TzPlane} for illustration.

Moreover there are other quantization conditions. Pick the compact cycle $B_i$ of the Riemann surface whose intersection number with $A_i$ is $\delta_{ij}$. The field equation of $T(z)$ implies that
\begin{equation}
\frac{1}{2\pi i} \oint_{B_i} T(z) dz = - N'_i  \in \mathbb{Z}.
\label{quantizeB}
\end{equation}
This is proved by computing the effective superpotential and studying its field equations; so this relation is on shell \cite{Cachazo:2003yc}. Quantization condition of the resolvent $T(z)$ over cycles of $\Sigma$ implies that $T(z) = d \log \xi(z)$ for some function $\xi(z)$ on Riemann surface $\Sigma$. 

Another way of understanding the quantization condition for $T(z)$ is as follows. Once we expand the anomaly equations \eqref{anomaly eqs} and impose \eqref{QCasN+p}, the set of equations are overdetermined; there are more equations than variables. In order for the recursion relation to admit solutions, it is necessary and sufficient that the periods of $T(z)$ are quantized. This statement is proved by Ferrari and collaborators \cite{Ferrari:2007bz, Ferrari:2007kk}. If one defines $T(z) = F'(z)/F(z)$ then \cite{Ferrari:2007kk} concludes that
\begin{equation}
F(z) + \frac{\gamma B(z)}{F(z)} = P(z)
\label{nonpertT}
\end{equation}
with degree $N$ polynomial $P(z)$. Then
\begin{equation}
F(z) = \frac{1}{2} \pbra{P(z) + \sqrt{P^2(z) - 4 \gamma B(z)}}, 
\label{T det}
\end{equation}
and 
\be
T(z) = \frac{d}{dz} \log \pbra{P(z) + \sqrt{P^2(z) - 4 \gamma B(z)}}.
\label{QuantumT}
\ee
The factor $\gamma$ can be chosen so that when $m(z) = (M + z) \delta^{\tilde f}_f$, in the square root of \eqref{QuantumT} $P(z)^2 - 4\gamma B(z)$ should reduce to standard Seiberg-Witten curve; when $m(z) = m^{\tilde f}_f$ it should reduce to that of \cite{Cachazo:2002ry}. Therefore it is natural that $\gamma = \Lambda^{2N - N_f}$. This is consistent with \cite{Cachazo:2003yc}. By setting $\Lambda = 0$ one can get back to the classical results: 

\begin{equation}
T(z) = \frac{F'(z)}{F(z)} = \frac{P'(z)}{P(z)}
\end{equation}
so the degree $N$ polynomial $P(z)$ can actually be identified as $\det (z - \Phi)$, that is why we used the same symbol as that of \eqref{clCasrelation}. Note that the expression in the square root of \eqref{QuantumT} is precisely what is conjectured by Kapustin \cite{Kapustin:1996nb} to be the $\CN = 1$ analogue of Seiberg-Witten curve. 

The quantum corrected formula \eqref{QuantumT} \textit{is} a chiral ring relation, since \eqref{QuantumT} is satisfied on all supersymmetric vacua of the theory.

The photino $w_{\alpha}$ will be corrected as well. \eqref{nonpertT} holds for arbitrary $\Phi$, so it holds for $\Phi + \epsilon M$ for arbitrary small $\epsilon$ and any matrix $M$. Taking derivative with respect to $\epsilon$ in $T(z) = F'(z)/F(z)$, we have
\be
\Tr \frac{M}{(z - \Phi)^2} = -\partial_z \pbra{\frac{F^{(\epsilon)}(z)}{F(z)}},
\label{PhotinoNP}
\ee
where we have introduced
\be
F^{(\epsilon)} = -\partial_{\epsilon} F(z; \epsilon) =-\partial_{\epsilon}\bbra{ \frac{1}{2} \pbra{P(z; \epsilon) + \sqrt{P^2(z; \epsilon) - 4 \gamma B(z)}}}
\ee
with $P(z; \epsilon) = \det(z - \Phi - \epsilon M)$. Take $M = W_{\alpha}$ and integrate over \eqref{PhotinoNP}, we get
\be
w_{\alpha} = \left. \frac{1}{4\pi} \frac{-\partial_{\epsilon} P(z; \epsilon)}{\sqrt{P^2(z) - 4 \gamma B(z)}} \right |_{\epsilon \rightarrow 0}.
\ee
This is a new relation. However, as $w_{\alpha}$ has trivial expectation value for supersymmetric vacua, we will not need this relation in the future.

\subsubsection{Comparison with perturbative chiral ring}

After nonperturbative analysis, let us take a quick look at how perturbative ring looks like. By perturbative chiral ring we mean the strong coupling scale $\Lambda \rightarrow 0$, and the chiral ring relation is governed by one-loop Konishi anomaly alone. 

First we show that perturbatively there is no gaugino condensations. Recall our theory is governed by Riemann surfaces parametrized by $y(z)^2 = {\widetilde W}'(z)^2 + f(z)$. The nonperturbative formula \eqref{QuantumT} gives another parametrization of the Riemann surface $\Sigma$: $P^2(z) - 4\Lambda^{2N-N_f} B(z)$. Requiring consistency of the theory means the Riemann surfaces must factorize properly \cite{Cachazo:2003yc}:
\be
P^2(z) - 4\Lambda^{2N-N_f} B(z) & = H^2(z) C(z),\\[0.5em]
{\widetilde W}'(z)^2 + f(z) & = G(z)^2 C(z),
\ee
where $G(z)$ and $H(z)$ are some polynomials. Perturbatively $\Lambda = 0$, so we see ${\widetilde W}'(z)^2 + f(z)$ is a perfect square. However since ${\widetilde W}'(z)$ has degree $k$ while $f(z)$ has degree $k-1$, this is impossible unless $f(z) = 0$. Plug into \eqref{quantumR}, we see we must have $R(z) = 0$. Plug into \eqref{anomalyM}, we go back to the classical $F$-term for the Higgs branch. Therefore, perturbation theory does not alter the classical Higgs branch vacua.

\subsection{Examples of chiral ring solution}\label{subsec:physical}

We have introduced the gadgets to compute the quantum chiral ring of the massive theory in previous subsections, $c.f.$ equations \eqref{anomaly eqs} and \eqref{QuantumT}. In this section we explicitly see how chiral ring solutions give supersymmetric quantum vacua, in a one-to-one manner. 

Let us consider a massive $U(2)$ theory with one flavor, and $k=2$. This model is considered in section \ref{subsec3.2}; here we assume the tree level superpotential to be
\begin{equation}
W_{\rm tree} = \frac{1}{3} {\rm Tr} \Phi^3 - \frac{1}{2} {\rm Tr} \Phi^2 + {\tilde Q}(1 + \Phi) Q,
\label{deformed212}
\end{equation}
where we pick all the coupling to be $\pm 1$ for simplicity. Let us focus first on classical chiral ring. The expectation value of $\Phi$ can have either pseudo-confining vacua or Higgs vacua (modulo Weyl equivalence):
\begin{equation}
\langle \Phi \rangle = \left(\begin{array}{cc} 0\ & 0 \\ 0\ & 0\\[0.2em]  \end{array} \right), \ \ \left( \begin{array}{cc} 0 & 0 \\ 0 & 1\\[0.2em]  \end{array} \right), \ \ \left( \begin{array}{cc} 1 & 0 \\ 0 & 1\\[0.2em]  \end{array} \right), \ \ \left( \begin{array}{cc} -1 & 0 \\ 0 & 0\\[0.2em] \end{array} \right), \left( \begin{array}{cc} -1 & 0 \\ 0 & 1\\[0.2em] \end{array} \right).
\label{clSol}
\end{equation}
This can be computed using entirely the chiral ring. Our strategy is to solve \eqref{anomaly eqs} and then rule out certain solution using $\eqref{6Anomaly}$. Classically there is no gaugino condensation so $R(z) = 0$. Expanding with respect to large $z$ we have 
\begin{equation}
\begin{aligned}
u_{n+2} - u_{n+1} + v_n & = 0, \\[0.5em]
v_{n+1}  +  v_n & = 0.
\end{aligned}
\end{equation}
These equations give $u_1 = u_3 = u_5 = \dots$, and $u_2 = u_4 = u_6 = \dots$. There are also chiral ring relations for the adjoints. We know from \eqref{clCasrelation}:
\begin{equation}
P(z) = \det (z - \Phi) = \sum_{i = 0}^N p_i z^{N-i}.
\end{equation}
The coefficients $p_i$ of $P(z)$ are related to $u_j$ by Newton's identity 
\begin{equation}
p_n = -\frac{1}{n} \sum_{i = 1}^n u_i p_{n-i},
\label{newton id}
\end{equation}
so we obtain two equations on the generators $u_1$ and $u_2$:
\begin{equation}
\begin{aligned}
u_2 - u_1^2 - \frac{u_2^2}{2} + \frac{u_1^2 u_2}{2} & = 0,\\[0.5em]
u_1 + \frac{u_1^3}{2} - \frac{3}{2} u_1 u_2 & = 0.
\end{aligned}
\end{equation}
These two equations actually contain six solutions, which are
\be
(u_1, u_2) = (0,0),\ (1,1),\ (2,2), \ (-1,1), \ (0,2), \ (-2,2).
\ee
Here the first five solutions are exactly listed in \eqref{clSol}, including both Coulomb and Higgs vacua; the last one is not a physical solution, which corresponds to putting two $-1$ (the root of $1+z$) in the diagonal of $\langle \Phi \rangle$. 

Remember that we still have one extra Konishi anomaly equation \eqref{6Anomaly}, which imposes additional constraint on generalized mesons. The recursion relation reads:
\be
(n+1) (v_{n+2} - v_{n+1}) + \sum_{i=0}^n v_i v_{n-i} = 0.
\ee
Notice that this equation is satisfied for all Coulomb branch vacua; the recurrence is also satisfied for the vacua $(u_1, u_2) = (-1,1)$ and $(0,2)$. However $(-2,2)$ is ruled out. Therefore, our classical chiral ring relation gives complete solution which is identical to solving the $F$-term equations.

In \cite{Cachazo:2003yc} the first five Konishi anomaly equations are used. There the way to make the solution physically sensible is to impose by hand that the residue of the resolvent $T(z)$ at the Higgs branch singularity should be at most $1$; this extra condition is valid both at classical and quantum level. We conjecture that this residue condition is equivalent as imposing another Konishi anomaly \eqref{6Anomaly}. We prove it in section \ref{subsec:onetoone}.

Next we would like to analyze the quantum chiral ring of the model \eqref{deformed212}. Quantum mechanically the anomaly equations read: 
\begin{equation}
\begin{aligned}
u_{n+2} - u_{n+1} + v_n & = 2 \sum_{i=0}^{n-1} r_i u_{n-i-1},\\[0.5em]
v_{n+1} + v_{n} & = r_n, \\[0.5em]
r_{n+2} - r_{n+1} & = \sum_{i=0}^{n-1} r_i r_{n-i-1},\\[0.5em]
(n+1) (v_{n+2} - v_{n+1}) + \sum_{i=0}^n v_i v_{n-i} & = 2 \sum_{i=0}^{n-1} (n-i) r_i v_{n-i-1}.
\end{aligned}
\end{equation}
Likewise we read off the constraints of Casimir operators by expanding
\begin{equation}
T(z) = \frac{P'(z)}{\sqrt{P^2(z)-4 \Lambda^3 (1+z)}} - \frac{2 \Lambda^3}{\sqrt{P^2(z)-4 \Lambda^3 (1+z)}} \frac{1}{P(z)+\sqrt{P^2(z)-4 \Lambda^3 (1+z)}}
\end{equation}
with $P(z) = p_0 z^2 + p_1 z + p_2$. Then we obtain the following relations on $u_i$:
\begin{equation}
\begin{aligned}
u_3 & = 3 \Lambda^3 -\frac{1}{2}u_1^3 + \frac{3}{2} u_1 u_2,\\[0.5em]
u_4 & = 4 \Lambda^3 (1+ 2 u_1) - \frac{1}{2} u_1^4 + u_1^2 u_2 + \frac{1}{2} u_2^2,\\[0.5em]
u_5 & = 10 \Lambda^3 \left(u_1^2 + u_1 + \frac{1}{2} u_2 \right)-\frac{1}{4} u_1^5 + \frac{5}{4} u_1 u_2^2,\\[0.5em]
u_6 & = 9 \Lambda ^6-\frac{3}{4} u_1^4 u_2+ 18 \Lambda ^3 \left( \frac{1}{3} u_1^3 + \frac{2}{3} u_1^2 + u_1 u_2 + \frac{1}{3} u_2 \right) +\frac{3}{2}u_1^2 u_2^2+\frac{1}{4}u_2^3,\\[0.5em]
       & \cdots
\end{aligned}
\end{equation}

To the order of $u_6$ we can completely determine the expectation value of $u_1$ and $u_2$ and get rid of any unphysical solutions. One can use the elimination theory to get the final equation for $u_1$:
\begin{equation}
\begin{aligned}
(u_1 - 1) \pbra{u_1^3 (u_1 + 1) (u_1-2)^2 - 9u_1 (8u_1^2 + 9 u_1 + 4) \Lambda^3 -27 \Lambda^6} = 0.
\end{aligned}
\end{equation}
Note that the vacua are corrected by instantons. When setting $\Lambda \rightarrow 0$ we get back to the classical solutions. In particular we recognize one vacuum in the solution with eigenvalue ${\rm diag}(0,1)$ for $\langle \Phi \rangle$. When $u_1 = 1$, we can solve that $u_2 = 1$, thus determine the characteristic polynomial $P(z) = z^2 - z$. For generalized glueballs we have $2 r_0 = r_1 = 2\Lambda^3$. Therefore we can package it as  
\be
T(z) & = \frac{d}{dz} \log \left[ z^2 - z + \sqrt{(z^2 - z)^2 -4 \Lambda^3 (1+z)} \right],\\[0.5em]
R(z) & = \frac{1}{2} \left( z^2 - z -\sqrt{(z^2 - z)^2 - 4 \Lambda^3(z-1) - 8 \Lambda^3} \right), \\[0.5em]
M(z) & = \frac{R(z)}{1+z}.
\ee
For this solution, the two Riemann surfaces defined by $y(z)^2 = {\widetilde W}'(z)^2 + f(z)$ and ${\tilde y}(z)^2 = P(z)^2 - 4\Lambda^3 B(z)$ match exactly. The reason that $u_1 = 1$ is not quantum corrected by instantons is that this vacua corresponds to residual $U(1) \times U(1)$ gauge symmetry; Coulomb branch vevs leave both monopoles massive, so in the low energy there are still two independent photons. Moreover from the expression of $T(z)$ we know in this case instanton corrections begin to enter only for superpotential with $k \geq 3$.

\textit{Isomorphism of Coulomb branch vacua.} In writing down the quantum chiral ring associated to \eqref{deformation}, we see that the only quantity that enters into the formula is $B(z) = \det \bbra{m^{\tilde f}_f(z)}$, which is a degree $lN_f$ polynomial. This means for various choices of $l$ and $N_f$, one can pick distinct $l$ and $N_f$ such that $B(z)$ is identical. It is natural to conjecture that for these choices the Coulomb branch vevs are exactly the same. This is confirmed by explicit examples (for one example, see appendix \ref{isoCoulomb}), thus prove the claim of \cite{Kapustin:1996nb}.

\subsection{Solution of the chiral ring and supersymmetric vacua}\label{subsec:onetoone}

We now turn to the proof that solutions of the chiral ring in the mass deformed theory are in one to one correspondence with supersymmetric vacua. We also show that, the extra anomaly equation \eqref{6Anomaly} implies residue constraint on the Higgs branch, proposed by \cite{Cachazo:2003yc}.

We begin by proving that the one-to-one correspondence holds for Coulomb branch vacua. Classically, it is obvious that those vacua are exactly contained in the chiral ring by setting $\langle Q \rangle = \langle \tilde Q \rangle = 0$ and $R(z) = 0$ in the Konishi anomaly \eqref{anomaly eqs}:
\be
{\rm Tr} \frac{{\widetilde W}'(\Phi)}{z - \Phi} = 0,
\ee
since this is just a gauge invariant way of writing $F$-term equations.

Conversely, we show the solution of Konishi anomaly is contained in $F$-term solution. For Coulomb branch vacua, the proof is very similar to that of \cite{Svrcek:2003az}. One can write
\be
0 = {\rm Tr} \frac{{\widetilde W}'(\Phi) - {\widetilde W}'(z) + {\widetilde W}'(z)}{z - \Phi} = -\zeta(z) + {\widetilde W}'(z) T(z),
\ee
where $\zeta(z)$ is a degree $k-1$ polynomial. Therefore we have an equality:
\be
T(z) = \frac{P'(z)}{P(z)} = \frac{\zeta(z)}{{\widetilde W}'(z)}, 
\ee
or in the product form $P'(z){\widetilde W}'(z) = \zeta(z) P(z)$. Over complex field $\mathbb{C}$ the polynomials can be factorized, so the general solution is of the form
\be
& \zeta(z) = E(z) {\tilde \zeta}(z), \ \ \ \ P(z) = F(z) H(z),\\[0.5em]
& {\widetilde W}'(z) = E(z) F(z), \ \ \ \ P'(z) = {\tilde \zeta}(z) H(z).
\ee
Then we have $T(z) = {\tilde \zeta(z)} / F(z)$. But the root of $F(z) = \prod_{i=1}^n (z - \lambda_i)$ is the subset of root of ${\widetilde W}'(z)$, and since ${\tilde \zeta}(z)$ is of degree $n-1$, so we obtain:
\be
T(z) = \sum_{i=1}^n \frac{\nu_i}{z- \lambda_i}.
\label{1to1Tz}
\ee
By definition of $T(z)$ one concludes that all $\nu_i$ are integers, labelling the number of entries of $\lambda_i$ in the diagonal of $\langle \Phi \rangle$. So this solution can be obtained by solving $F$-term.

Next we turn to the classical Higgs branch. This part of the proof is new. Again it is obvious that the $F$-term equations admit solutions that are all solutions of chiral ring relations. Conversely, suppose the fractional decomposition of resolvent $T(z)$ is:
\be
T(z) = \sum_I \frac{r_I}{z-b_I} + \dots,
\ee
where the dots represent the terms coming from roots of ${\widetilde W}'(z)$ as in \eqref{1to1Tz}. Moreover we also claim \cite{Cachazo:2003yc} the solution of $M(z)$ classically is given by \eqref{Mcl}. Plug into \eqref{anomalyM} we examine the singular part in $z$ while ignore regular part and obtain:
\be
m^{\tilde f}_f (z) M^f_{\tilde f}(z) = 0.
\ee
We integrate this formula around $b_I$ and noticing the singularity comes from $M(z)$ while $m(z)$ is a polynomial, we conclude that $m^{\tilde f}_f (b_I)$ is a degenerate matrix, namely
\be
B(b_I) = \det m^{\tilde f}_f (b_I) = 0,
\ee
so $b$ must be a root of $B(z)$. However, straightforward computation shows that the Konishi anomaly equations \eqref{anomalyC} - \eqref{anomalytM} even admits solution of $T(z)$ and $M(z)$ with $r_I > 1$. This is exactly what happens in section \ref{subsec:physical}. We now show that the sixth anomaly equation \eqref{6Anomaly} imposes the condition $r_I = 0$ or $1$.

For simplicity and avoiding clutter of notation, we assume the superpotential to be $W_Q = m_1 {\tilde Q}Q + m_2 {\tilde Q}\Phi Q$ but we keep ${\widetilde W}_{\Phi}$ generic. Moreover, to linearize \eqref{6Anomaly} we restrict our chiral rotation to be
\be
\delta \Phi = \frac{1}{z - \Phi} {\tilde Q}_{\tilde g} h^{\tilde g}_g Q^g,
\ee
then it is not hard to see the singular part of \eqref{6Anomaly} becomes
\be
W'(z) M(z)_{\tilde g}^g + v_{0, {\tilde g}}^f m_{2,f}^{\tilde f} M(z)_{\tilde f}^g= 0
\ee
with $M(z)$ being substituted with explicit expression we arrive at $-r_I + r_I^2 = 0$, namely it can only take value $0$ and $1$. In proving this we have used the following fact:
\be
\frac{1}{(2 \pi i)^2} \oint_{b_I} \oint_{b_J} \pbra{ \frac{1}{m(x)}}^f_{\tilde g} m^{\tilde f}_{2, f} \pbra{ \frac{1}{m(y)}}^g_{\tilde f} dx dy = \frac{\delta_{IJ}}{2\pi i} \oint_{b_I} \pbra{\frac{1}{m(x)}}^g_{\tilde g} dx.
\ee

The conclusion with $r_I = 0,1$ is exactly the same as the residue condition proposed in \cite{Cachazo:2003yc}. Therefore we conclude that the solution of chiral ring is in one to one correspondence with the supersymmetric vacua at classical level. 

We now comment on the correspondence at the quantum level. We again divide our vacua into Coulomb branch and Higgs branch. Note first that the residue condition $r_I = 0, 1$ cannot be modified at the quantum level. Otherwise if one turns off the strong coupling scale $\Lambda$ and perturbative anomaly, then the residue condition at classical level is violated. Put another way, an integral constraint is robust against quantum corrections.

On the Coulomb branch, the low energy behavior is determined by factorization of the matrix model curve $y(z)^2 = {\widetilde W}'(z)^2 + f(z)$. If there are $k-n$ massless monopoles, then we have
\be
{\widetilde W}'(z)^2 + f(z) & = H_{k-n}^2(z) F(z), \\[0.5em]
P(z)^2 - 4 \Lambda^{2N_c - N_f} B(z) & = Q_{N-n}^2(z) F(z),
\ee
so that $F(z)$ is a degree $2n$ polynomial, giving a genus $n-1$ Riemann surface. The number of independent photinos is $n$. The period of the resolvent $T(z)$ around cycles of Riemann surface give the unbroken rank of the gauge group. These vacua degenerates in a one-to-one manner to the classical supersymmetric vacua.

\subsection{Massless limit and Kutasov model}\label{subsec:Massless}

We have seen how to calculate the classical and quantum chiral ring of the mass deformed theory by means of solving the recursion relations. In this subsection we will approach the massless limit, by setting
\be
g_n \rightarrow 0, \ \ \ (n < k) \ \ \ \ \ \ \text{and} \ \ \ \ \ m^{\tilde f}_f(z) \rightarrow 0,
\ee
and obtain the moduli space of vacua for massless Kutasov model. Again, we emphasize that in this way we only recover the radical of the ring relations as an ideal.

\subsubsection{How many parameters are enough?}\label{subsec:No.parameter}

Unlike ordinary SQCD \cite{Seiberg:1994bz} where ${\Tr} m M$ is the only choice of single trace operator deformation, for Kutasov model there are many more deformation parameters. Just as written in \eqref{deformation}, we may add

\begin{enumerate}
\item[(1)] Casimir deformations:  $g_n {\Tr} \Phi^{n+1}$ for $n < k$;
\item[(2)] Generalized meson deformations:  ${\Tr} m_n v_n$ for $n N_f < 2N_c$ \cite{Cachazo:2003yc,Kapustin:1996nb}\footnote{The reason for this requirement is that (1) the generalized meson deformations are all relevant; (2) the metric of the Coulomb branch is positive definite; (3) the ${\CN=2}$ theory whose curve is isomorphic to that in the square root of \eqref{QuantumT} is asymptotically free.}.
\end{enumerate}

Generally, it is required to add all deformations and take various allowed limits. Unfortunately, it would be a cumbersome task. We would like to examine their physical significance and whether their number could be reduced.

Let us begin by Casimir deformations, \eqref{deformPhi}. For $k > 1$, these deformations are used to resolve the nilpotent matrix $\Phi$ into a semisimple matrix, $c.f.$ section \ref{subsec:generalities}.  Let there be $s_1$ of $a_1$ in the diagonal of $\langle \Phi \rangle$. The low energy gauge group contains a factor $U(s_1)$ and some $W$-bosons become massive, with mass 
\be
M_{W} = \left | a_1 - a_i \right|, \ \ \ \ i \neq 1,
\ee
and $\Phi$ acquires mass which is function of $a_i$'s as well. So tuning $g_i$'s is essentially tuning physical mass parameters. Therefore we have to at least keep the mass generic and distinct; hence the most general \eqref{deformPhi} is required.

Next we turn to generalized meson deformation \eqref{deformQ} with $l N_f < 2N_c$. The claim is that \textit{if one takes generic limit\footnote{By generic limit we mean that the roots of ${\tilde W}'(z)$ and $B(z)$ are kept distinct.}, only the first meson deformation, $\Tr m v_0$ is sufficient}. We expect such limit probes a subset of true quantum moduli space.

To understand this, we compare the most general deformation \eqref{deformQ} and deformation using only ${\Tr} m v_0 = m_f^{\tilde f} {\tilde Q}_{\tilde f}Q^f$. It is quite obvious that two cases share identical Coulomb branch vacua. For the latter, there is no Higgs branch vacua classically; while for the former case, it is given by \eqref{classicalHiggs}. 

Now we take the generic limit. From \eqref{HiggsQ} we learn that the second term in left hand side approaches to a finite quantity while the terms in the bracket goes to zero as $m^{\tilde f}_{f,n} \rightarrow 0$. To have solutions we must require at least one of $\langle Q^f_1 \rangle$ and $\langle {\tilde Q}_{\tilde f}^1 \rangle$ goes to infinity, which is a run-away vacua. Therefore, we conclude that the extra Higgs branch vacua are absent; the two kinds of deformation are equivalent.

Quantum mechanically, the solution of $M(z)$ for arbitrary vacuum is given by \cite{Cachazo:2003yc}:
\begin{equation}
M(z) = R(z) \frac{1}{m(z)} - \sum_{I=1}^L \frac{r_I {\widetilde W}'(z_I) + (1-2 r_I)R(q_I)}{z-z_I} \frac{1}{2\pi i} \oint_{z_I} \frac{1}{m(x)} dx
\end{equation}
where $z_I$ for $I = 1, \dots L = l N_f$ is the roots of $B(z)$. When $r_I = 1$, poles of $T(z)$ around $z_I$ is on the first sheet, while $r_I = 0$ the second sheet. When all $r_I = 0$, we return to the Coulomb branch vacua, \eqref{quantumM}. A fact that we will prove in the Appendix \ref{subsec:propChiral} is $R(z) = 0$ in the final limit, so if there exists some $r_J = 1$ we see that the second term of $M(z)$ is infinite, assuming no accidental cancellation appears.

However, in the classical expression \eqref{HiggsQ}, we see a flat direction opens up if $b$ happens to be the root of ${\widetilde W}'(z)$. These would recover some missing Higgs branches. Therefore, to completely reproduce the flat directions in the quantum vacua, $B(z) = \det \bbra{m^{\tilde f}_f (z)}$ should have at least many roots as ${\widetilde W}'(z)$. Therefore, we conjecture the sufficient number of meson deformations should satisfy:
\be
k - 1 \leq lN_f < 2N_c.
\label{mesonBound}
\ee
Here we write $k - 1$ instead of $k$, as an overall $U(1)$ factor in the gauge group does not affect the result.


Even for $l = 1$, the computation of chiral ring is quite challenging. Relegating detailed study for future, here we only focus on the potential with $l = 0$:
\be
W_{\rm tree} = \sum_{n=0}^k \frac{g_n}{n+1} {\rm Tr} \Phi^{n+1} +  m_f^{\tilde f} {\tilde Q}_{\tilde f}Q^f
\label{minimalW}
\ee
to probe a subset of vacuum structure. We will see in certain cases it already has very nontrivial consequences. For convenience, we will take $g_k = 1$ in later examples. Note that $k=2$ is special. We know the most general deformation is ${\tilde W}'(z) = z^2 + \theta z +\nu$. No matter which root one picks, we always get the mass
\be
\left | {\tilde W}''(z_{1,2}) \right | = \Delta_2 = \sqrt{\theta^2 - 4\nu}
\ee
so what matters is the discriminant. We can thus set $\theta = 0$ for a further simplification.



With the deformation ${\Tr} m v_0$ only, the six Konishi anomaly equations are no longer mutually independent. In fact, the anomaly \eqref{6Anomaly} can be deduced from \eqref{anomalyr} and \eqref{anomalyM}. We have seen that this is true classically in section \ref{subsec:physical}. Quantum mechanically we can expand \eqref{6Anomaly} in terms of $z \rightarrow \infty$:
\be
(n+1) \sum_{i=0}^k g_i v_{n+i} = 2 \sum_{i=0}^{n-1} (n-i) r_i v_{n-i-1},
\ee
where we omitted the flavor indices. Now multiplying both sides by mass matrix $m^{\tilde g}_g$, using \eqref{anomalyM}, and massage the dummy indices a little we get
\be
2 (n+1) \sum_{i=0}^k g_i r_{n+i} = 2(n+1)  \sum_{i=0}^{n-1} r_i r_{n-i-1}.
\ee
We see this is exactly the recursion relation given by $\eqref{anomalyr}$. Therefore in the following computation we will ignore \eqref{6Anomaly} unless stated.

In Appendix \ref{subsec:propChiral}, we examine some general properties of the vacuum expectation values in the massless limit, from the recursion relations.



\section{Examples of quantum chiral rings}\label{sec5:examples}

In this section we study examples of massless $\hat S$. These various examples also give further confirmation on the statements we made previously, $e.g.$ section \ref{subsec:No.parameter} and at the beginning of section \ref{sec4:quantum}.

\subsection{$k=1$: the vacua of $U(N_c)$ SQCD}

We begin with $k=1$, the superpotential \eqref{superpotentialX} is essentially a mass term. When the scale $\Lambda$ of the theory is smaller than the mass scale of the adjoint, $\Phi$ can be integrated out in the IR and the theory is effectively given by $U(N_c)$ SQCD. This RG flow has been analyzed in \cite{Argyres:1996eh,Bolognesi:2008sw}, while $U(N_c)$ SQCD was studied in \cite{Chen:2011wn,Shifman:2007kd}. Since $\Phi$ is invisible in the IR, there is no need to add Casimir deformation \eqref{deformPhi}, in consistent with \eqref{mesonBound}.  

$U(N_c)$ SQCD with $N_f$ fundamental flavors can be thought of as gauging the baryon symmetry of $SU(N_c)$ theory, under which the quark and anti-quark have charge $\pm 1$ respectively. When $N_c \geq N_f$ the classical chiral ring is generated by mesons ${\tilde Q}_{\tilde f} Q^f$ freely; while for $N_c < N_f$ there are nontrivial relations among mesons \cite{Chen:2011wn}:
\be
M_{j_1}^{\left[ i_1\right.} M_{j_2}^{i_2} \dots M_{j_{N_c+1}}^{\left. i_{N_c+1}\right]} = 0.
\ee
This relation arises since mesons of order $N_f$ are built from rank $N_c$ data; and in particular for $N_f = N_c+1$ the relation becomes $\det M = 0$. 

In the following we scale the mass $g_1$ in \eqref{superpotentialX} to be $1$, and its dependence can be easily recovered. The superpotential we use is
\be
W_{\rm tree} = \frac{1}{2} {\rm Tr} \Phi^2 + m^{\tilde f}_f {\tilde Q}_{\tilde f} Q^f.
\ee

A short cut to analyze the quantum vacua is to directly apply \eqref{CSW T}. However we will try a more elaborated way by solving the recursion relation directly. This will be helpful later.

First, the recursion relation for generalized glueball in \eqref{anomalyr} can be solved explicitly:
\begin{equation}
r_{n+1} = \sum_{i=0}^{n-1} r_i r_{n-i-1},
\end{equation}
which is actually a recursion relation for binomial coefficients in $(1+x)^{1/2}$. By induction,
\begin{equation}
r_{2j}  = \frac{2^j (2j-1)!!}{(j+1)!} r_0^{j+1}, \ \ \ \ r_{2j+1} = 0.
\end{equation}

Next we focus on the recursion relation \eqref{anomalyC}:
\begin{equation}
u_{n+1} = 2 \sum_{i=0}^{n-1} r_i u_{n-i-1}.
\end{equation}
Similar induction tells us that
\begin{equation}
u_{2j}  = \frac{2^j (2j-1)!!}{j!} r_0^{j}u_0, \ \ \ \ u_{2j+1} = 0, 
\end{equation}
with initial condition $u_0 = N_c$. We can plug them into the series of $T(z)$ and get
\begin{equation}
\begin{aligned}
T(z) & = \sum_{n=0}^{+\infty} \frac{u_{2n}}{z^{2n+1}} \\[0.5em]
       & = \frac{u_0}{z} \sum_{n=0}^{+\infty} \left( \begin{array}{c} -\frac{1}{2}\\ n \end{array} \right) \left( -\frac{4r_0}{z^2} \right)^n \\[0.5em]
       & = \frac{N_c}{z} \pbra{1-\frac{4r_0}{z^2}}^{-\frac{1}{2}}
\end{aligned}
\end{equation}
which is exactly the same as given by \eqref{CSW T}. 

We define ${\tilde \Lambda}^{2N_c} = (\det m) \Lambda^{2N_c-N_f}$ by scale matching condition. In the meanwhile that there is a degree $N_c$-polynomial $P(z)$ with leading coefficient $1$ such that
\begin{equation}
T(z) = \frac{N_c}{z} \pbra{1-\frac{4r_0}{z^2}}^{-\frac{1}{2}} = \frac{P'(z)}{\sqrt{P(z)^2 - 4 {\tilde \Lambda}^{2N_c}}}.
\end{equation}
Integrating both sides and note that the only way that $P(z)$ is a polynomial with leading coefficient $1$ is that
\begin{equation}
r_0 \sim {\tilde \Lambda}^2 = (\det m)^{\frac{1}{N_c}} \Lambda^{\frac{2N_c - N_f}{N_c}}.
\label{SQCDr}
\end{equation}
Hence,
\begin{equation}
\langle {\tilde Q}Q \rangle  = v_0 = \pbra{\frac{1}{m}} r_0 = \Lambda^{\frac{2N_c - N_f}{N_c}} (\det m)^{\frac{1}{N_c}} \pbra{\frac{1}{m}}.
\label{SQCDM}
\end{equation}
However, we should remember the scale $\Lambda$ appeared here is not the scale $\Lambda_L$ of low energy effective SQCD. They are related by scale matching condition
\be
\Lambda_L^{3N_c-N_f} = \Lambda^{2N_c-N_f}.
\ee
Substituting into \eqref{SQCDr} and \eqref{SQCDM} we see the results for vevs of mesons and gaugino condensation is exactly given by that of \cite{Seiberg:1994pq,Seiberg:1994bz}.

Now we can list the quantum chiral ring for above cases. 
\begin{enumerate}
\item[(1)] $N_c > N_f$. There is no supersymmetric ground state; which means the ideal $\hat \CS$ contains unit, so $\hat \CR$ is empty;
\item[(2)] $N_c = N_f$. It is easy to see $\det v_0 = \Lambda^{N_c}$. Therefore the quantum moduli space is smoothed out. 
\item[(3)] $N_c < N_f$. The quantum moduli space is the same as the classical one, thus
\be
{\hat \CR}_{N_c, N_f, 1} = \CR_{N_c, N_f, 1}.
\ee
\end{enumerate}

\subsection{$U(2)$ theory with $k = 2$ revisited}

In this section we analyze the quantum chiral ring of the examples given in \ref{subsec3.2}. As mentioned before we will use the superpotential \eqref{minimalW} to deform the Kutasov model:
\be
W_{\rm tree} = \frac{1}{3} {\rm Tr} \Phi^3 - \tau^2 {\rm Tr} \Phi + m^{\tilde f}_f {\tilde Q}_{\tilde f} Q^f,
\label{U(2)k2W}
\ee 
where we define $\tau^2 = -g_0$. We can use \eqref{anomalyC} and \eqref{anomalyr} to solve for the Casimir $u_j$ and generalized glueball $r_j$ first. There are two types of solution:

\vspace{5pt}
$\bullet$ $1^{st}$ Solution\footnote{In writing solution like this, we assume the convention $(x)^{\frac{1}{2}} = \pm \sqrt{x}$, namely one can flip \textit{simultaneously} the sign for the square root. So the above solution has in fact four independent solutions. We do not apply this rule to the strong coupling scale $x = \Lambda$.}:
\be
u_1 & = - \bbra{4\tau^2 - 8(\det m)^{\frac{1}{2}} \Lambda^{\frac{4-N_f}{2}}}^{\frac{1}{2}},\\[0.5em]
r_0 & = - \pbra{\det m}^{\frac{1}{2}}\, \Lambda^{\frac{4-N_f}{2}} \bbra{4\tau^2 - 8(\det m)^{\frac{1}{2}} \Lambda^{\frac{4-N_f}{2}}}^{\frac{1}{2}},\\[0.5em]
r_1 & = \pbra{\det m}^{\frac{1}{2}}\, \Lambda^{\frac{4-N_f}{2}} \left[ 2\tau^2 - 3(\det m)^{\frac{1}{2}} \Lambda^{\frac{4-N_f}{2}}\right].
\label{22Sol1}
\ee 

$\bullet$ $2^{nd}$ Solution:
\be
u_1 & = 0, \ \ \ u_2 & = 2 \tau^2,\ \ \ r_0 & = 0,\ \ \ r_1 & = (\det m) \Lambda^{4 - N_f}.
\ee
Higher order operators are zero in the limit. These two solutions are in fact the quantum deformed version of the classical vacua $[1,1]$ and $[2]$ in section \ref{subsec3.2}. Indeed, the classical Coulomb vacua for the massive theory is either ${\rm diag}(\tau, \tau)$ or ${\rm diag}(\tau, -\tau)$. The corresponding Young tableau is $[\dl{2}]$ and $[\dl{1},\dl{1}]$, which is dual to the Young tableau of nilpotent matrix $[1,1]$ and $[2]$. 

The vevs of generalized meson is related to glueballs by $\eqref{anomalyM}$ as $v_j = r_j m^{-1}$. The resulting quotient is an indeterminate, whose value depend on how $\tau$ and $m_f^{\tilde f}$ approach to zero. 

\vspace{5pt}
$\bullet$ $N_f = 1$. For the vacuum $[2]$, we see in the massless limit:
\be
u_1 = 0, \ \ 
u_2 = 0, \ \ 
v_0 = 0, \ \ 
v_1 = \Lambda^3.
\ee
This vacuum is quantum mechanically modified, as we have $kN_f = N_c$. Going back to table \ref{chargeTab}, we see immediately that the charge of $\Lambda^3$ is exactly the same as the charge of $v_1$. This is consistent with holomorphy. However, we fail to produce flat direction for $v_0$ in this particular limit.

For the vacua $[1,1]$ we see that
\bse
\begin{align}
v_0 & = - 2 m^{-\frac{1}{2}} \Lambda^{\frac{3}{2}} \bbra{\tau^2 - 2 m^{\frac{1}{2}} \Lambda^{\frac{3}{2}}}\label{212v0}^{\frac{1}{2}},\\[0.5em]
v_1 & = m^{-\frac{1}{2}} \Lambda^{\frac{3}{2}} \left[ 2\tau^2 - 3m^{\frac{1}{2}} \Lambda^{\frac{3}{2}} \right].
\end{align}
\ese 
Here we have the freedom to tune parameters $\tau$ and $m$ simultaneously. Consider
\begin{equation}
\tau^2 - 2m^{\frac{1}{2}}\Lambda^{\frac{3}{2}} \approx \eta^2 m^{\alpha} \Lambda.
\end{equation}
For \eqref{212v0} not to diverge in the limit, we must have $\alpha \geq 1$. To the leading order we may pick $\alpha =1$. Plug in, we see
\be
v_0  = -\eta \Lambda^2 \in \mathbb{C}, \ \ v_1 = \Lambda^3,
\ee
so $v_1$ is again corrected by one-instanton effect, although it has zero classical moduli. We conjecture $v_1 = \Lambda^3$ holds for the entire vacua from all possible limit. Note because of this the Higgs branch of Kutasov model is smoothed out, there are no singularities on the moduli space. This is the $k = 2$ analogue of smooth moduli space for $N_c = N_f$ in SQCD.

Here we see a qualitative difference between Kutasov model and its deformed cousin. If we keep the deformation parameter $\tau$ finite, then taking $m \rightarrow 0$ gives divergent $v_0$ and $v_1$. This is in accordance with \cite{Kutasov:1995ve,Kutasov:1995ss}; the finite $\tau$ endows adjoint chiral multiplet a mass, so the low energy effective theory is just $U(2)$ SQCD with one flavor. It is a well-known fact that ADS superpotential lift the vacuum and the theory does not have a ground state \cite{Affleck:1983mk,Seiberg:1994pq,Seiberg:1994bz}. But this will not happen in Kutasov model where we have seen that simultaneous parameter-tuning still preserves the flat direction.

\vspace{5pt}
$\bullet$ $N_f = 2$. This is the simplest case when the theory is in conformal window \cite{Kutasov:2003iy,Gukov:2015qea}. Now $m_f^{\tilde f}$ is a $2\times 2$ matrix. For simplicity we will take it to be diagonal, $m = {\rm diag}(\mu_1, \mu_2)$.

Consider vacuum $[2]$ first. Everything remains the same except there is no instanton correction anymore: $v_1 = 0$. For vacuum $[1,1]$, the expressions are similar:
\bse
\begin{align}
v_0 & = - 2\pbra{\det m}^{\frac{1}{2}}\, \Lambda \bbra{\tau^2 - 2(\det m)^{\frac{1}{2}} \Lambda}^{\frac{1}{2}} \pbra{\frac{1}{m}},\\[0.5em]
v_1 & = \pbra{\det m}^{\frac{1}{2}}\, \Lambda \left[2\tau^2 - 3(\det m)^{\frac{1}{2}} \Lambda \right] \pbra{\frac{1}{m}}.
\end{align}
\ese
We see no matter how one tunes the parameter, $v_1$ is always zero in the limit\footnote{For instance, we can consider the tuning
\begin{equation}
\tau^2 - 2(\mu_1 \mu_2)^{\frac{1}{2}}\Lambda \approx \eta \mu_1^{\alpha} \mu_2^{\beta} \Lambda^2, \ \ 0 < \alpha, \beta < 1
\end{equation}
where we choose $\alpha, \beta < 1$ for the reason that $v_0$ does not diverge. One sees that 
\begin{equation}
v_1 \propto \mu_1^{\frac{\alpha}{2}}  \mu_2^{\frac{\beta}{2}} v_0
\end{equation}
after dropping factors which is zero in the limit. Since $\mu_{1,2} \rightarrow 0$ and $\alpha, \beta$ are positive, we see $v_1 \rightarrow 0$ in the limit.}. We conclude that generic massless limit could not recover flat directions for $v_1$. However, it is possible to give flat direction to $v_0$.

The origin of Higgs branch $v_0 = 0$ remains. This means that at the singularity, the $SU(2)_L \times SU(2)_R$ chiral symmetry is unbroken, and the theory is in non-abelian Coulomb phase. The IR behavior exhibits Kutasov duality.

Here we can also see the difference between Kutasov model and its deformed cousin. When $\tau$ is finite, we have $\det v_0 = 4\tau^2 \Lambda^2$. Since the adjoint superfield $\Phi$ is massive with mass $2\tau$, we see $4\tau^2 \Lambda^2$ is nothing but the low energy scale $\Lambda^4_L$ of SQCD. This is precisely the quantum modified moduli space of SQCD.

\subsection{$U(3)$ theory with $k=2$ revisited}

Next we turn to the $U(3)$ theory whose classical chiral ring is analyzed in section \ref{subsec:classicalU(3)}. The superpotential deformation used is again \eqref{U(2)k2W}.

$\bullet$ $1^{st}$ Solution. This is the one corresponding to $[2,1]$ vacuum:

\be
& u_1 = -\tau, \ \ \ u_2 = 3\tau^2,\\[0.5em]
& r_0 = - \pbra{\det m}^\frac{1}{2} \Lambda^{\frac{6 - N_f}{2}},\\[0.5em]
& r_1 = \pbra{\det m}^\frac{1}{2} \Lambda^{\frac{6 - N_f}{2}} \tau.
\ee

$\bullet$ $2^{nd}$ Solution. This is the one corresponding to $[1,1,1]$ vacuum:
\be
& u_1 = -3\sqrt{\tau^2 - 2 (\Lambda^{6-N_f}\det m)^{\frac{1}{3}}},\\[0.5em]
& u_2 = 3\tau^2, \\ 
& r_ 0 = -2 (\Lambda^{6-N_f}\det m)^{\frac{1}{3}} \sqrt{\tau^2 - 2 (\Lambda^{6-N_f}\det m)^{\frac{1}{3}}},\\[0.5em]
& r_1 = 2 (\Lambda^{6-N_f}\det m)^{\frac{1}{3}} \bbra{\tau^2 - \frac{3}{2} (\Lambda^{6-N_f}\det m)^{\frac{1}{3}}}.
\ee
To get the vevs of generalized mesons $v_0$ and $v_1$ we again divide $r_0$ and $r_1$ by mass matrix $m$. 

We mainly focus on $N_f = 1$ and this is the region for $kN_f < N_c$. We immediately see $[2,1]$ vacua is non-existent. For $[1,1,1]$ vacuum, we have to be more careful since there is a possibility of tuning parameters. However, to make $v_0$ finite we need to set:
\be
\tau^2 - 2 (\Lambda^{6-N_f}\det m)^{\frac{1}{3}} \propto m^{\frac{4}{3}} + \text{higher order terms}.
\ee
But this makes $v_1$ divergent. Therefore, the vacua is quantum mechanically erased, and the chiral ring is empty:
\be
{\hat R}_{3,1,2} = \varnothing.
\ee
This is consistent with the semi-classical analysis of \cite{Kutasov:1995np, Kutasov:1995ss}.

\subsection{Chiral ring relation from magnetic dual}

In \cite{Kutasov:1995ss}, Kutasov, Schwimmer and Seiberg conjectured a quantum chiral ring relation for the Casimir operators ${\Tr}\Phi^n$. Classically these operators are constrained by the superpotential terms as well as the characteristic polynomial of $\Phi$; however quantum mechanically the characteristic polynomial coming from the adjoint $\Psi$ in the magnetic theory should also be added to the electric theory, via duality maps that sends ${\Tr}\Psi^n$ to the combination of ${\Tr \Phi}^m$. In this way the quantum Coulomb vacua on both sides match.

Here we would like to check this statement explicitly. We consider $U(4)$ theory with $N_f = 3$ and $k=2$ with mass deformation only for adjoint field $\Phi$:
\be
W_{\Phi} = \frac{1}{3} {\rm Tr}\Phi^3 - \frac{1}{2} {\rm Tr}\Phi^2.
\ee
Classically, the theory has five vacua that are labelled by diagonal entries of $\langle \Phi \rangle = {\rm diag}(0,0,0,0)$, ${\rm diag}(0,0,0,1)$, ${\rm diag}(0,0,1,1)$, ${\rm diag}(0,1,1,1)$, ${\rm diag}(1,1,1,1)$. This can be packaged into two equations obtained from characteristic polynomial as follows. From Konishi anomaly equation \eqref{anomaly eqs}, we set the right hand side of \eqref{anomalyC} to zero and get the recursion relation:
\be
u_{n+2} - u_{n+1} = 0.
\ee
Moreover, the fact that $T(z) = P'(z) / P(z)$ where $P(z)$ is a degree $4$ polynomial implies that $u_i$ for $i > 4$ can be expressed by $u_{1,2,3,4}$. Using above recursion relation we can easily obtain:
\be
u_2 \pbra{u_2^4 - 10u_2^3 + 35 u_2^2 - 50u_2 + 24} = 0,
\ee
which is the classical relation coming from ``electric" characteristic polynomial\footnote{Our results are slightly different from that of \cite{Kutasov:1995ss} in the sense that there are more vacua because the gauge group is unitary. For special unitary gauge group the traceless condition reduces the number of allowed vacua by about one half. Therefore, we would have $N_c / 2$ when $N_c$ is even as in \cite{Kutasov:1995ss}. }.

Let us now see what happens quantum mechanically. To compute quantum corrections we endow all quarks with mass by deforming the superpotential as\footnote{Because $\Phi$ is massive now, deforming by $\Tr m v_0$ is enough.} 
\be
W = \frac{1}{3} {\rm Tr}\Phi^3 - \frac{1}{2} {\rm Tr}\Phi^2 + m_f^{\tilde f}{\tilde Q}_{\tilde f}Q^f
\ee
and we expect some of the vacua would be erased when $m_f^{\tilde f} \rightarrow 0$. Indeed such vacuum has two types of solutions. For the first one, it is a deformation of $\langle \Phi \rangle = {\rm diag} (0,0,0,0)$:
\be
u_1 & = 2 - 2\bbra{1 - 8 \pbra{\det m \Lambda^5}^{\frac{1}{4}}}^{\frac{1}{2}},\\[0.5em]
v_0 & = - (\det m \Lambda^5)^{\frac{1}{4}} \pbra{\frac{1}{m}}.
\ee
Moreover, since $N_f = 3$ we learned that $\det v_0$ is infinite. Therefore this vacuum is not present at quantum level. Similar reasoning shows that the vacuum which is the deformation of $\langle \Phi \rangle = {\rm diag} (1,1,1,1)$ is also absent. The total number is reduced from $5$ to $3$, corresponding to $u_1 = 1,2,3$.

Physically, these two run-away vacua precisely correspond to the parameter regime where ADS superpotential is generated at low energies ($N_f < N_c$) after $\Phi$ is integrated out. The idea of \cite{Kutasov:1995ss} is that such elimination is equivalent to including the characteristic polynomial from magnetic dual via operator mapping. We now demonstrate that this is true.

First of all, it is straightforward to check that as $m_f^{\tilde f} \rightarrow 0$ the vevs of Casimir operators are not quantum shifted. Following \cite{Kutasov:1995ss} we define
\be
{\hat \Phi} = \Phi - \frac{1}{2} \mathbb{I},
\ee
where $\mathbb{I}$ is the unit matrix. Then the superpotential becomes:
\be
W_{\Phi} = \frac{1}{3}{\rm Tr} {\hat \Phi}^3 - \frac{1}{4}{\rm Tr} {\hat \Phi} - \frac{1}{3}.
\ee

Kutasov duality proposes that the magnetic dual is a $U(2)$ gauge theory with $N_f = 3$ flavors of quarks and generalized mesons, plus an adjoint field $\Psi$ with superpotential
\be
{\hat W} = {\hat W}_{\Psi} + {\hat W}_{q} = \sum_{i=0}^2 \frac{{\hat g}_i}{i+1} {\rm Tr} \Psi^i + \sum_{j=0}^1v_j {\tilde q}\ \Psi^{1-j} q.
\ee
When focusing on Coulomb branch, we can perform a similar trick and turn the superpotential of $\Psi$ part into
\be
{\hat W}_{\Psi} = \frac{{\hat t}_0}{3} {\rm Tr} {\hat \Psi}^3 + {\hat t}_2 {\rm Tr} {\hat \Psi} + {\hat \alpha}
\ee
where ${\hat \alpha}$ is some constant. The coupling and operator mappings given in \cite{Kutasov:1995ss} tell us that
\be
{\hat t}_0 = 1, \ \ \ \ {\hat t}_2 = -\frac{1}{4}.
\ee
Then it is not hard to see that for dual theory, the Coulomb branch has three allowed choices:
\be
\langle {\hat \Psi} \rangle = \left( \begin{array}{cc} 1/2 & 0  \\ 0 & 1/2\\[0.3em] \end{array} \right), \ \ \ \left( \begin{array}{cc} -1/2 & 0  \\ 0 & 1/2\\[0.3em] \end{array} \right), \ \ \ \ \left( \begin{array}{cc} -1/2 & 0  \\ 0 & -1/2\\[0.3em] \end{array} \right),
\ee
and one can deduce the magnetic characteristic polynomial following the same procedure as before:
\be
{\hat u}_1^3 - {\hat u}_1 = 0, \ \ \ \ {\hat u}_2 = \frac{1}{2}
\ee
with ${\hat u}_i = {\rm Tr} {\hat \Psi}^i$. Applying the operator mapping derived in \cite{Kutasov:1995ss} we have
\be
{\rm Tr} {\hat \Psi} = - {\rm Tr} {\hat \Phi} = -{\rm Tr} \Phi + 2,
\ee
so we need to add to the electric theory one more constraint, which is
\be
0 & = (-u_1 + 2)^3 - (-u_1+2) \\[0.5em]
   & = - u_1^3 + 6u_1^2 - 11 u_1 + 6,
\ee
the solution of which is restricted to $u_1 = 1,2,3$, exactly as that computed directly from chiral rings of electric theory.

\acknowledgments{This work is funded by the DOE Grant DE-SC0011632 and the Walter Burke Institute for Theoretical Physics. The author is especially grateful to Sergei Gukov, Ken Intriligator, Emily Nardoni, Yu Nakayama for their patient explanation and various stimulating discussions throughout the project. Moreover KY would also like to thank Noppadol Mekareeya, Satoshi Nawata, Du Pei, Pavel Putrov and Dan Xie for helpful comments.}

\appendix

\section{General properties of the recursion relations}\label{subsec:propChiral}

In this appendix we wish to extract some universal properties of the vacua for all $N_c, N_f$ and $k$ with deformation \eqref{minimalW}, and the massless limit.

Before diving into technical proof, we may imagine how the vacuum looks like by physical argument. First, we know $\Phi$ is classically nilpotent, labelled by a set of discrete integers. In other words, $\Phi$ is already ``quantized" at the classical level, and quantum corrections cannot modify it. So we expect $u_j = 0$ quantum mechanically as well. Moreover, the superpotential \eqref{superpotentialX} truncates the chiral ring, and we expect this is also true at quantum level. Specifically, we expect there exists an integer $k_0$ such that for $j \geq k_0$ all $v_j = 0$. Classically $k_0 = k$.

We prove the following claims. Some claims can be proven even for most general deformations \eqref{deformation}. We will use a * notation to indicate this situation.

\vspace{8pt}
\textit{Claim 1*. All generalized glueball has trivial vevs $r_j = 0$, implying $R(z) = 0$. Thus $U(N_c)$ Kutasov model does not have non-trivial gaugino condensations.}

\textit{Proof.} From Konishi anomaly \eqref{anomalyM}, we can expand around $z \rightarrow +\infty$ and look at coefficients of $z^{-n-1}$. It reads: 
\begin{equation}
\sum_{j=1}^{l+1} m_{f,j}^{\tilde f} v^{f'}_{{\tilde f}, n+j-1} = \delta^{f'}_{f} r_n.
\end{equation} 

A physically sensible solution of the quantum chiral ring should have all the elements $u_n$, $r_n$ and $v_n$ as functions of parameters $\{g_i, m_{f,l}^{\tilde f} \}$, and they must be finite when the parameters approach zero. Therefore taking the limit of both sides of above equations, and picking $f = f'$, we immediately see
\begin{equation}
r_n = 0.
\end{equation}
In particular, $r_0 \propto {\rm Tr}W_{\alpha}W^{\alpha} \sim \langle \lambda \lambda \rangle = 0$. \hfill $\Box$

\vspace{8pt}

\textit{Claim 2. There exists $k_0$ such that for all $j \geq k_0$, $v_j = 0$ in the chiral ring.} 

\textit{Proof}. Here we assume superpotential \eqref{minimalW}. Then Konishi anomaly \eqref{anomalyM} and \eqref{anomalytM} tell us that
\begin{equation}
[m.v_n]^{f'}_f = \delta^{f'}_f r_n, \ \ \ [v_n . m]^{\tilde f}_{{\tilde f}'} = \delta^{\tilde f}_{{\tilde f}'} r_n,
\end{equation}
which means $m$ and $v_n$ commute and the product is a diagonal matrix, proportional to $r_n$ times the identity. Then from the Konishi anomaly \eqref{anomalyr}
\begin{equation}
\sum_{i=0}^k g_i r_{n+i}  = \sum_{i=0}^{n-1} r_i r_{n-i-1}.
\end{equation}

One can think of it as a matrix equation, and substitute each $r_n$ by $m.v_n$ and multiply $m^{-1}$ on both sides. Taking limit on both sides we see $v_{k+n} = 0$ for all $n \geq 0$. Thus the truncation is at least as far as in classical case. \hfill  $\Box$

\vspace{8pt}
\textit{Claim 3*. $u_{k+n} = 0$ for all $n \geq 0$.}

\textit{Proof.} This time we use Konishi anomaly \eqref{anomalyC}. One obtains
\begin{equation}
\sum_{i=k}^0 g_i u_{n+i} + \sum_{j=1}^{l+1} (j-1) m_{f,j}^{\tilde f} v^{f}_{{\tilde f}, n+j-2} = 2 \sum_{i=0}^{n-1} r_i u_{n-i-1}.
\end{equation}

Again taking the limit on both sides and use the condition that $r_n = 0$ of claim 1, and all parameters except $g_k$ is infinitesimally small, we see that $u_{k+n} = 0$ for any non-negative integer $n$. \hfill $\Box$

\vspace{8pt}

\textit{Claim 4. $u_1 = u_2 = \dots = u_{k-1} = 0$.}

\textit{Proof.} We will use induction. Notice first that
\be
T(z)^2 \pbra{P(z)^2 - {\tilde \Lambda}^{2N}} = P'(z)^2
\ee
where ${\tilde \Lambda}^{2N} = (\det m) \Lambda^{2N-N_f}$ and $P(z) = p_N + p_{N-1} z + \dots p_1 z^{N-1} + z^N$. It is now safe to take massless limits on both side\footnote{Here one should first show that $p_i$ are all finite in the limit. Indeed, with deformation \eqref{minimalW} $p_i$ can be expressed by polynomial of $u_1, \dots u_N$ and no instanton factor would enter. In other words the expressions are the same as classical case.}, and because of claim 3, we obtain an equality:
\be
\pbra{\frac{u_{k-1}}{z^k} + \dots + \frac{N}{z}} & \pbra{p_N + p_{N-1} z + \dots +p_1 z^{N-1} + z^N} \\[0.5em]
& = p_{N-1} + 2 p_{N-2} z + \dots + (N-1) p_1 z^{N-2} + N z^{N-1}
\label{zeroCasimir}
\ee
Now suppose $k = 2$. The comparing coefficients on both sides tells us $u_1 p_N = 0$. Then we must have $p_N = 0$, otherwise we are done. Then by iterating the procedure we see $p_1 = p_2 = \dots = p_N = 0$; then $u_1 = 0$ so the claim is valid for $k=2$.
Suppose the claim is true for $k-1$, now we proceed to the case of $k$.  Again by comparing the coefficients of \eqref{zeroCasimir}, under the condition $u_{k-1} \neq 0$ (otherwise we are done by assumption), we see all $p_i$'s vanish. Therefore, $u_{k-1}$ must vanish as well. So the proof is complete. \hfill $\Box$

Although expectation values of Casimir operators and generalized mesons are zero, they may not be trivial in the chiral ring. We conclude that quantum mechanically, in general the chiral ring of Kutasov model can still be written as
\be
{\hat {\CR}}_{N_c,N_f,k} = \mathbb{C}[u_1, u_2, \dots, u_{k-1}, v_0, v_1, \dots, v_{k-1}] / {\hat \CS}(u_1, u_2, \dots, u_{k-1}, v_0, v_1, \dots, v_{k-1}),
\ee
where we have omitted the generalized glueball and photinos $\w_{\alpha, k}$.

\section{Isomorphism of Coulomb branch vacua}\label{isoCoulomb}

In this appendix, we consider two examples that the quantum Coulomb branch receive exactly the same corrections. We take the gauge group to be $U(2)$.

\subsection{$N_f = 1$, $l = 2$}

We pick the superpotential to be
\be
W = \frac{1}{3} \Tr \Phi^3 - \frac{1}{2} \Tr \Phi^2 + {\td Q}(2 + 3\Phi + \Phi^2) Q.
\ee
The recursion relation becomes:
\be
u_{n+2} - u_{n+1} + 2 v_{n+1} + 3 v_{n} & = 2 \sum_{i = 0}^{n-1} r_i u_{n-i-1},\\[0.5em]
r_{n+2} - r_{n+1} & = \sum_{i = 0}^{n-1} r_i r_{n-i-1},\\[0.5em]
 v_{n+2} + 3 v_{n+1} + 2 v_n & = r_n,\\[0.5em]
(n+1) (v_{n+2} - v_{n+1}) + 3 \sum_{i = 0}^n v_i  v_{n-i} + & 2 \sum_{i = 0}^{n} v_i  v_{n-i+1} = 2 \sum_{i=0}^{n-1} (n-i) r_i v_{n-i-1}.
\label{Recur:U(2)Nf1k2l2}
\ee

\textit{Classical vacua}. At classical level one can set the right hand side of above recurrence formulae to be zero and only consider the first, third and fourth equations. Then one can first solve the generalized mesons:
\be
v_n = (-2)^n C_1 + (-1)^n C_2
\ee
where $C_{1,2}$ are two parameters that determine the initial condition. Then we can further plug the expression in the first equation of \eqref{Recur:U(2)Nf1k2l2} and eliminate additional variables. So the classical chiral ring relation for $u_1$ is
\be
(u_1 - 2)(u_1 - 1) u_1 (u_1 + 1)(u_1 + 2)(u_1 + 3) = 0 
\ee
This precisely corresponds to $3$ Coulomb branch vacua and $5$ Higgs branch vacua.

\textit{Quantum vacua}. The quantum recursion relation can be solved leaving single generator $u_1$ as usual. We expect that the quantum moduli space is a deformation of the classical one in the sense that if we take the strong coupling scale $\Lambda \rightarrow 0$, we should recover classical chiral ring, possibly with increased multiplicities of the roots. Indeed in this case we have
\be
 (u_1 - 1)(u_1 + 3) & \left ( u_1^8 - (7 + 52 \Lambda^3) u_1^6 - (2 + 376 \Lambda^3) u_1^5 + (12 - 926 \Lambda^3 - 204 \Lambda^6) u_1^4 \right. \\[0.5em] & \left. + (8 - 1000 \Lambda^3 - 976 \Lambda^6) u_1^3 - (498\Lambda^3 + 1552 \Lambda^6 + 160 \Lambda^9) u_1^2 \right. \\[0.5em] & \left. - (100\Lambda^3 + 1120 \Lambda^6 + 448 \Lambda^9) u_1 -275 \Lambda^6 - 160 \Lambda^9 + 64 \Lambda^{12} \right) = 0
\label{QuantumU2k2Nf1l2}
\ee

\subsection{$N_f = 2$, $l = 1$}

We take the superpotential to be
\be
W = \frac{1}{3} \Tr \Phi^3 - \frac{1}{2} \Tr \Phi^2 + m^{\td f}_{1,f} {\td Q}_{\td f} Q^f + m^{\td f}_{2,f} {\td Q}_{\td f} \Phi Q^f,
\ee
and we use the chiral symmetry to cast $m_1$ into diagonal form and is assumed to be
\be
m^{\td f}_{1,f} = \left( \begin{array}{cc} 1 & 0 \\ 0 & 2 \\[0.5em]\end{array} \right),
\ee
while in principle $m_2$ does not have to be diagonal, but we require it to be invertible. To make things simple we set
\be
m^{\td f}_{2,f} = \left( \begin{array}{cc} 1 & 0 \\ 0 & 1 \\[0.5em]
\end{array} \right).
\ee

\textit{Classical vacua}. The recursion relation is
\be
& u_{n+2} - u_{n+1} + \Tr\, m_2 . v_n = 0,\\[0.5em]
& m_{1,f}^{\td f} v_{n, \td f}^{f'} + m_{2,f}^{\td f} v_{n+1, \td f}^{f'} = 0,\\[0.5em]
& (n+1) (v_{n+2} - v_{n+1})_{\td g}^g + \sum_{i=0}^n v_{i, {\td g}}^f m_{2, f}^{\td f} v_{n-i, {\td f}}^g = 0.
\ee
From the second equation we see $v_{n} = - m_1 . v_{n-1} = (-m_1)^n v_0 = v_0 (-m_1)^n$. This fact means $v_0$ must be a diagonal matrix, so are all generalized mesons. Then one can again eliminate variables and obtain the relation for the generator $u_1$, so that we arrive at
\be
 (u_1 - 2)(u_1 - 1) u_1 (u_1+ 1)(u_1 + 2)(u_1 + 3) = 0,
\ee 
and also the recursion relation could uniquely determine the vevs of generalized mesons.

\textit{Quantum vacua}. The right hand side of recursion relations should be supplemented by the anomalies. Since one also has $v_n . m_1 = m_1 . v_n$ so generalized mesons are still diagonal. The nonperturbative corrections to Casimir operators:
\be
T(z) = \frac{d}{dz} \log \pbra{P(z)^2 + \sqrt{P(z)^2 - 4 (1+z)(2+z) \Lambda^2}}
\ee
is in fact the same as $N_f = 1, l = 2$ case, except the substitution $\Lambda^3 \rightarrow \Lambda^2$. After some lengthy calculation we obtain the relation for the generator $u_1$:
\be
 (u_1 - 1)(u_1 + 3)& \left ( u_1^8 - (7 + 52 \Lambda^2) u_1^6 - (2 + 376 \Lambda^2) u_1^5 + (12 - 926 \Lambda^2 - 204 \Lambda^4) u_1^4 \right. \\[0.5em] & \left. + (8 - 1000 \Lambda^2 - 976 \Lambda^4) u_1^3 - (498\Lambda^2 + 1552 \Lambda^4 + 160 \Lambda^6) u_1^2 \right. \\[0.5em] & \left. - (100\Lambda^2 + 1120 \Lambda^4 + 448 \Lambda^6) u_1 -275 \Lambda^4 - 160 \Lambda^6 + 64 \Lambda^{8} \right) = 0.
\ee
It is not surprising to see that the expression is isomorphic to \eqref{QuantumU2k2Nf1l2}, and the quantum shift to the chiral ring generator $u_1$ is exactly the same. This isomorphism can be attribute to the fact that the curve $P(z)^2 - 4 \Lambda^{2N_c - N_f} B(z)$ is isomorphic on the Coulomb branch.

\newpage

\bibliographystyle{JHEP_TD}
\bibliography{draft}

\end{document}